\documentclass[pre,aps,twocolumn,showpacs,floatfix]{revtex4}
\usepackage{amsmath,amssymb,graphicx}
\usepackage{epsfig}
\usepackage{textcomp}  
\usepackage{graphicx}
\usepackage[caption=false]{subfig}
\begin{document}

\title{Parameters inference and model reduction for the Single-Particle Model of Li ion cells.}
\author{M. Khasin$^1$,  C. S. Kulkarni$^1$ and K. Goebel$^2$}

\address{$^1$ SGT Inc., NASA Ames Research Center, Moffett Blvd, Mountain View, CA 94035, \\ $^2$ Palo Alto Research Center, Palo Alto CA 94304, Lulea University of Technology, Lulea, Sweden}

%\ead{michael.khasin@nasa.gov}

\begin{abstract}
The Single-Particle Model (SPM) of Li ion cell \cite{Santhanagopalan06, Guo2011} is a computationally efficient and fairly accurate model for simulating Li ion cell cycling behavior at weak to moderate currents. The model depends on a large number of parameters describing the geometry and material properties of a cell components. In order to use the SPM for simulation of a 18650 LP battery cycling behavior, we fitted the values of the model parameters to a cycling data. We found that the distribution of parametric values for which the SPM fits the data accurately is strongly delocalized in the (nondimensionalized) parametric space, with variances in certain directions larger by many orders of magnitude than in other directions. 

This property of the SPM  is known to be shared by a multitude of the so-called "sloppy models" \cite{Brown2003, Waterfall2006}, characterized by a few stiff directions in the parametric space, in which the predicted behavior varies significantly, and a number of sloppy directions in which the behavior doesn't change appreciably. As a consequence, only stiff parameters of the SPM can be inferred with a fair degree of certainty and these are the parameters which determine the cycling behavior of the battery. Based on geometrical insights from the Sloppy Models theory, we derive an hierarchy of reduced models for the SPM. The fully reduced model depends on only three stiff effective parameters which can be used for the battery state of health characterization.

%The set of parametric values for which the SPM prediction is consistent with the cycling data can be pictured as an effective multi-dimensional manifold in the parametric space, which we term Best-Fit Manifold (BFM) for brevity. Moving on the BFM toward limiting values of the sloppy parameters (e.g., zero or infinity) allows reduction of the SPM model to a model with fewer sloppy parameters with an insignificant decrease of fidelity. We derive an hierarchy of reduced models for the SPM. The fully reduced model depends on only three stiff effective parameters which can be used for the battery state of health characterization. 
\end{abstract}

\maketitle

\section{Introduction} \label{sec:SPMintro}
Li ion batteries have become increasingly important power supply in a broad range applications, from cells-phones to all electric cars to  UAVs and satellites. In the future it is expected that their use in aerospace applications will increase even more. Efficient performance monitoring  depends in particular on efficient and accurate models.

Data-driven models may be used to predict the cycling behavior of a battery, but lack interpretability in physical terms and require large amount of data to excel. Physics-based models of cycling behavior can incorporate prior knowledge of the battery as a physical system, require significantly less data for fitting and allow interpretation of the data in physical terms, facilitating development of physics-based models of aging and degradation.

A multitude of physics-based models have been developed, varying in accuracy, efficiency and breadth of validity \cite{Rahn13}. However, as expected, accuracy is generally traded for efficiency. 
High fidelity models are accurate for a wide range of cycling rates but are too computationally expensive to allow for efficient computation required in prognostic applications. 
The Single-Particle model (SPM) of an electrode was proposed in Ref.{\cite{Haran98}} and extended to Li ion systems and thermal behavior in Refs.\cite{Ning2004, Santhanagopalan06, Guo2011}. It is a computationally efficient model which neglects effects of the concentration and voltage variations in electrolyte, thereby allowing representation of each electrode as a single particle. Its range of validity is limited to low to moderate cycling rates and sufficiently thin cell architectures.

In view of its efficiency and satisfactory accuracy in the range of validity, we chose the SPM as a candidate physics-based model for prognostic applications in the field of UAVs. We had expected the implementation of the model to follow the following steps:
1. Fitting parameters of the model to available cycling data, providing a set of their best-fit values;
2. Using the best-fit values to predict the short-term cycling behavior of the battery;
3. Using the best-fit values to characterize  the battery's physical state - the state of health (SOH), and its long-term evolution - aging and degradation.

We found that parameter inference from the SPM of cycling behavior is an ill-posed problem: rather than providing a set of best-fit parametric values with reasonably narrow confidence intervals, fitting the SPM to cycling data delivers a continuum of parametric values, such that uncertainties in certain directions of the parametric space are of the order of the physical scale of allowed parameters variation. Uncertainties are small in only few directions of the parametric space, which, remarkably, do not all coincide with the original parameters of the SPM. As a consequence, a set of "best-fit" parametric values, which allows an accurate prediction of the cycling behavior, does not generally have a physical meaning, in the sense that it does not characterize the physical state of the battery. 

These properties of the SPM are shared by models of a multitude of complex systems  in various fields ranging from  system biology to radioactive decay (see the references below), where the focus of a model is limited to an emergent behavior \footnote{The term \textit{emergent behavior} is applied to an observed dynamics of complex (multi-component) system, which is an outcome of the interaction of the system's parts and cannot be reduced to effects of noninteracting components.} of the complex system. The corresponding models belong to the so-called Sloppy Models  universality class  \cite{Waterfall2006}. The research of the sloppy models over the past decade \cite{Brown2003, Waterfall2006, Gutenkunst2007, Transtrum2010, Transtrum2011, Machta2013, Transtrum2014,  OLeary2015, Niksic2016, Raman2017, Bohner2017, Raju2018, Mattingly2018} has provided theoretical insights and practical methods which enable consistent characterization of  complex systems. The main result of the theory is that only a few "stiff" effective parameters of a sloppy model affect its predictions and therefore, can be inferred from the data. The remaining independent effective parameters are "sloppy" - they do not affect the predictions and cannot be inferred. This property of the model can be interpreted as a statement about the complex system: just a small number of effective parameters of the system affects its emergent behavior.  To the best of our knowledge, the theory has never been applied to engineering systems, even though one can expect that dynamical models of complex engineering systems belong to the Sloppy Models universality class. 

Dealing with a sloppy model poses two main questions:  i) what are effective "stiff" parameters of the system and ii) how to infer them from the available data using the model. These two questions can be addressed using the so-called Manifold Boundary approximation method, introduced in Ref. \cite{Transtrum2014}, which provides a systematic way of reducing the original model to a model depending on just stiff parameters. Fitting the reduced model to the data, one can infer the values of the stiff parameters and characterize the state of the system. This approach has been applied to a number of complex systems ranging from system biology \cite{Transtrum2014} and biophysics \cite{Bohner2017} to nuclear physics \cite{Niksic2016}. 

In simpler cases, the reduction can be performed based on geometrical insights drawn from a numerical analysis of stiff and sloppy directions in the distribution of the best-fit values in the parameters space. This simpler approach is taken in the present work.
We build a hierarchy of reduced models starting with the SPM for predicting the cycling performance of the battery. The reduction allows  characterization of the battery physical state in terms of just three stiff parameters. One of the stiff parameters is a nonlinear function of the original  parameters of the SPM and does not have a "microscopic" interpretation.  This observation has important implications for modeling the battery aging and degradation based on the cycling data alone. 

In section II we present a short derivation and summary of the SPM using notation which will be useful later. Section III describes the adopted procedure for fitting the model parameters  and presents the results of the procedure: a distribution of parametric values in the parameter space consistent with the data. To make sense of this distribution, we review the theory of sloppy models in Section IV. Based on the geometrical insights of the theory and a numerical analysis we  develop a hierarchy of reduced models for the SPM  and offer their interpretation in terms of effective stiff parameters of a Li ion battery, Section V. Section VI is the concluding section.

\section{SPM}
\subsection{Single particle electrode assumptions.}
\paragraph{Spatial uniformity of electrolyte properties: concentration and voltage}
For a range of low and moderate currents, which depends on the electrolyte properties and the cell geometry, the variation of electrostatic potential in the electrolyte can be neglected, compared to potential drops on the solid/electrolyte interface  and Ohmic losses, the concentrations variations can be neglected with respect to the average ion concentration. Once the spatial variations  are  neglected, all the particles in the electrode become  equivalent with respect to their boundary conditions.  
\paragraph{ Uniformity of solid phase potential} It is assumed that solid phase is at uniform potential, which is  usually good approximation due to very large electronic conductivity of the electrodes.
\paragraph{Shape and size of the electrode particles}
The SPM assumes spherical shape of the solid electrodes particles and the same size for all the particles in a given electrode.
Although morphology of real electrodes do not justify these assumptions they are made due to a radical simplification of the model.

Based on the assumptions \textit{a-c} above, the SPM replaces all the particles in each electrode  by a single particle with the ion fluxes down-scaled by the total number of particles.

\paragraph{Constant solid diffusivity} Diffusivity of $Li$ in solid phase of electrodes is assumed to be independent of concentration. Experimental data does not generally support this assumption. However, data on solid phase diffusivity is not self-consistent, either, due to challenges in its measurement and interpretation of the results. For example, the Li diffusivity in graphite has been reported as both non-monotonous \cite{levi1997}, and constant  \cite{Yu1999}. From the theoretical perspective, solid diffusion of ions in the solid particle is an activated process and is related to the properties of the activity of ions \cite{FergusonBazant2012}. The latter is expressed in the properties of the open circuit voltage (OCV). Therefore, a consistent model should relate the two characteristic of the battery as noted in Ref.\cite{FergusonBazant2012}. However, for simplicity, the solid diffusivity is often assumed constant and the characteristics of the OCV are obtained as empirical correlations based on the data. This approach is taken by the SPM.

\subsection{Governing equations. Summary}
Dynamical equations for the SPM are derived in Appendix \ref{sec:SPM_eqs}.
The resulting expression for the time-dependent potential $V(t)$ are given below:
\begin{eqnarray} \label{eq:governing}
&&V(t)=\Delta \phi^{eq}_c(\bar{\theta}_c  )-\Delta \phi^{eq}_a(\bar{\theta}_a  )-I_a \left(r_a+r_c\right) \nonumber\\
&&+\frac{2 k_B T}{e} \ln\left(\frac{\chi_c  +\sqrt{(\chi_c  )^2+1}}{\chi_a  +\sqrt{(\chi_a  )^2+1}}\right),\\
&&  \chi_i  (\bar{\theta}_i)=\frac{I_i}{\tilde{I}_i \left({\theta}_i  \right)^{ \frac{1}{2}} \left(1-{\theta}_i  \right)^{ \frac{1}{2}}},\quad \tilde{I}_i\equiv  \frac{3 A l_i (1-\epsilon_i)k_i c_e^{ \frac{1}{2}}}{R_i}, \nonumber \\
&& I_a=-I_c, \ \ |I_i|=I, \ \ I_a > 0 \ \ for \ \ discharge, \nonumber \\
&& {\theta}_a=\bar{\theta}_a \Theta_{a,0}, \ \  {\theta}_c=1-\bar{\theta}_c\left(1- \Theta_{c,0}\right), \nonumber \\
&&Parameters \  (5):\  r=r_a+r_c, \ \tilde{I}_i, \ \Theta_{i,0} ; \  i=a,c, \nonumber
\end{eqnarray}
where  $r_i$ is Ohmic resistance, $\tilde{I}_i$ is specific exchange current and $\Theta_{i,0}$ is initial filling fraction of electrode $i=a,c$ and
\begin{eqnarray} 
&&\bar{\theta}_i(t)=1-{\mathcal{L}}^{-1}\left[\frac{\text{sign}(I_a){\mathcal{L}}\left[\bar{\zeta}_i( t')\right] }{  \sqrt{s} \coth {\sqrt{s}}-1}\right] \left(\frac{t}{\tau_i}\right), \label{eq:solbSPM2_sum} \\
&&\bar{\zeta}_i(t/\tau_i)= \frac{I(t) \tau_i }{3 e N_{Li,i}}, \nonumber \\ 
&&N_{Li,a}={M_{Li,a}\Theta_{a,0}}, \quad N_{Li,c}={M_{Li,c}\left(1-\Theta_{c,0}\right)}, \nonumber \\
&&\quad M_{Li,i} \equiv \frac{4}{3} \pi R_i^3K_i  m_i, \quad \tau_i\equiv\frac{R_i^2}{D_i}. \nonumber\\
&&Parameters \ (4):\  \tau_i, \ N_{Li,i}; \  i=a,c. \nonumber
\end{eqnarray}
where  $\tau_i$ is solid phase diffusion time and $N_{Li,i}$ is total number Li ions ($i=a$) or vacancies ($i=c$) of electrode $i=a,c$.

\section{Fitting the SPM to battery cycling data} \label{sec:fitting}
\begin{figure}[htp]
\vspace{-0.2cm}
\begin{center}
\includegraphics[width=3.3 in]{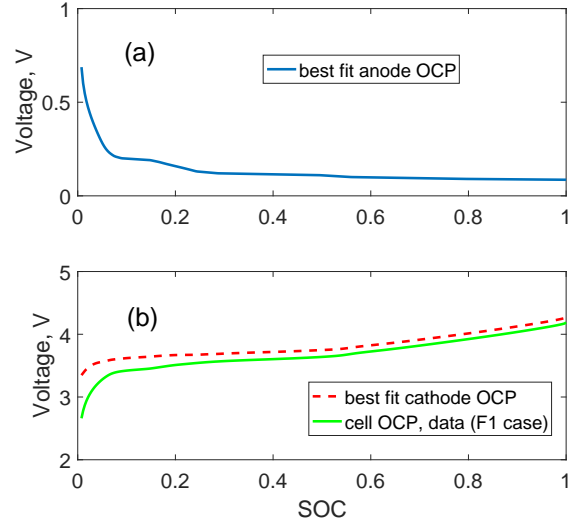} %the 0 in the beginning of the name should be eliminated
\caption{Subplot (a): Anode OCP, theory of Ref.\cite{Verbrugge2003} best fit to the cell OCP as explained in the main text. Subplot (b): Resulting cathode OCP (red dashed line) and cell's measured OCP.}
\label{fig:OCP_fit}
\end{center}
\end{figure}

As shown in the previous section the SPM is based on 9 parameters and two open circuit potentials (OCPs) - for the anode and the cathode. Both the parameters and the OCPs have to be fitted to the data. A common approach to fit the OCPs for cells with graphite anodes is to take the graphite anode OCP from literature, assuming that it is not different from the anode OCP in the tested cell, and to fit the cathode OCP using the measured total OCP. 
A problem with this approach is that graphite anode OCPs differ significantly from  anode to  anode leading. As a consequence, the inferred cathode OCPs will differ as well  and will generally display nonphysical features, inherited from the superposition of the measured total OCP and the assumed anode OCP. In contrast to graphite anodes which display a number of sharp transitions with the state of charge (SOC), OCP of the $LiCO_2$-based cathodes  are generally smooth \cite{Megahed94}. Therefore, the anode OCP should account for the sharp features observed in the measured OCP. This consideration has been a starting point for our approach to fitting of the anode and cathode OCP. To find the correct functional form for the anode OCP we used the theory of Ref. \cite{Verbrugge2003} to parametrize the anode OCP by 6 parameters which are then fit to the measured total OCP to give a maximally smooth cathode OCP. The details are given in Appendix \ref{app:OCP}. We note that the choice of the best fit parameters is not unique. The plot of the resulting anode and cathode OCPs for an early-life 18650 LP cell (case $F1$ in what follows) is given in Figure \ref{fig:OCP_fit}. 
\begin{figure}[htp]
\vspace{-0.2cm}
\begin{center}
\includegraphics[width=3.3 in]{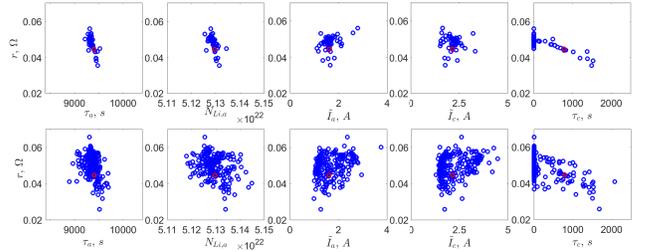} %the 0 in the beginning of the name should be eliminated 
\caption{Upper row: projections of an ensemble of best-fit parametric values, corresponding to the RMSE $\epsilon \le 1.02 \epsilon_{min|tr}$ for an application of Nelder-Mead minimization algorithm without an explicit stopping criterion; $\epsilon_{min|tr}\approx 2mV$ (red dot), i.e., $2\%$ of the total voltage drop during the discharge. Lower row: projections of an ensemble  $\epsilon \le 1.02 \epsilon_{min|tr}$ with a stopping criterion; $\epsilon_{min|tr}\approx 2mV$ (red dot), i.e., $2\%$ of the total voltage drop during the discharge.}
\label{fig:BFM20h8b_presentation2}
\end{center}
\end{figure}
The OCPs obtained using the described approach were kept fixed and used for the subsequent estimation of the 9 remaining parameters to the cycling data, which is the focus of the present work. To estimate the remaining parameters the standard least squares fitting approach was used. The best fit quality was assessed by the Root Mean Square Error (RMSE) of the model prediction compared to the data. The data used for parameters estimation was a set of discharge curves for various constant discharge currents within the range of validity of the SPM; e.g., $2A, 1A$ and $0.055A$ discharge currents were used. An initial guess for the values of the nine parameters was drawn from uniform prior distribution within a physically motivated boundaries, Appendix \ref{app:hyperparameters}. Next, the Nelder-Mead algorithm was used to search for the minimal value of the RMSE in the $9$-D parametric space, constrained by the boundaries, starting from the initial guess.
\begin{figure}[htp]
\vspace{-0.2cm}
\begin{center}
\includegraphics[width=3.3 in]{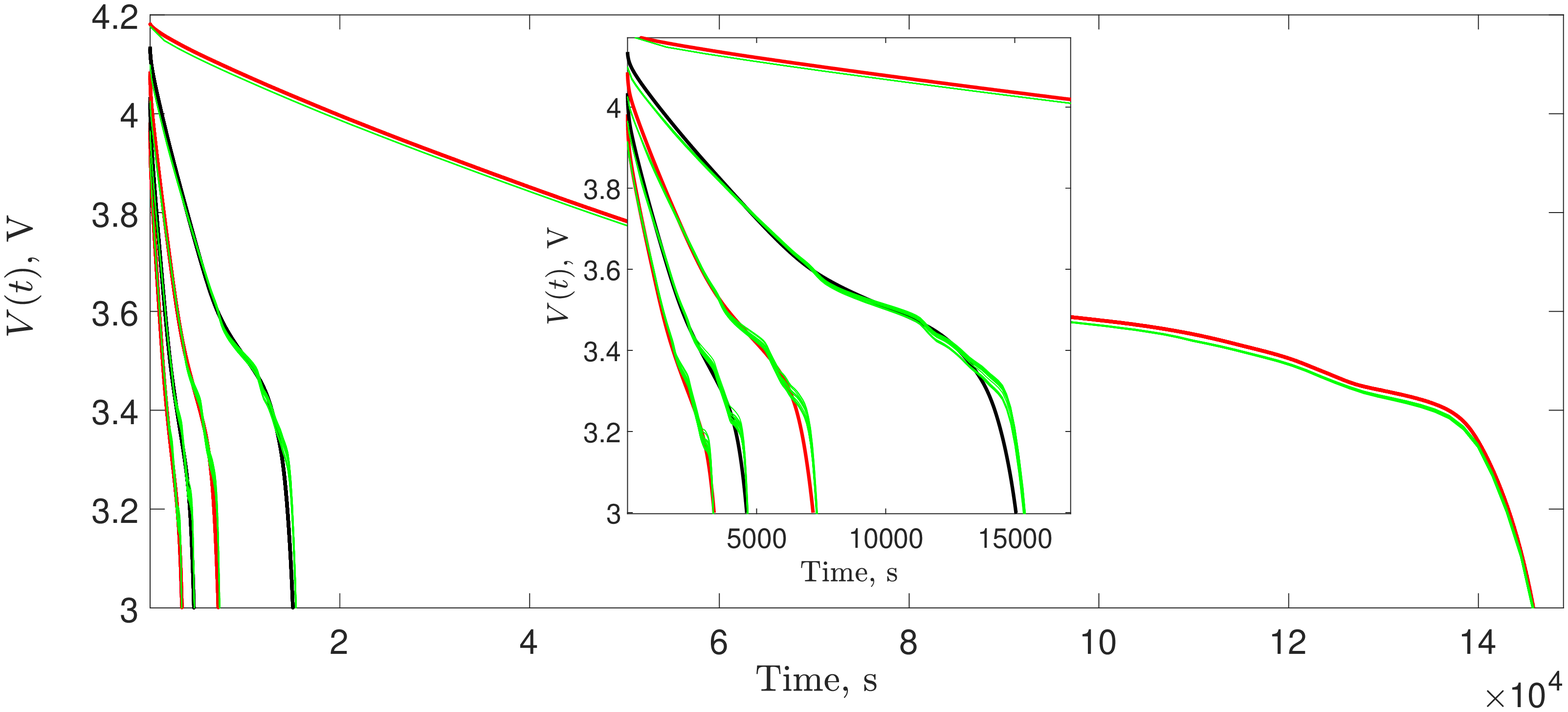} %the 0 in the beginning of the name should be eliminated pulse1ans2
\caption{Training (red) and testing (black) data vs model (green) for discharge currents: $2.0,1.5,1.0,0.5,0.055A$. 
The training data set is generated by $2.0,1.0,0.055A$ discharge currents. Twenty models' predictions (trajectories) are made based on $20$ random sets of parametric values within the range RMSE $\epsilon<1.02 \epsilon_{min|tr}$ - the minimal RMSE obtained for the training data; $\epsilon_{min|tr} \approx 2mV$. RMSE averaged over all the data is $2.2\%$ for each parametric set (the variation between the sets are negligible); average (over the data) error in the time of the end of discharge $t_{EOD}$ prediction is $1.15\%$(the variation between the sets are negligible). Computation time per trajectory: $0.1s$.}
\label{fig:dataVsFit}
\end{center}
\end{figure}
\begin{figure}[htp]
\vspace{-0.2cm}
\begin{center}
\includegraphics[width=3.3 in]{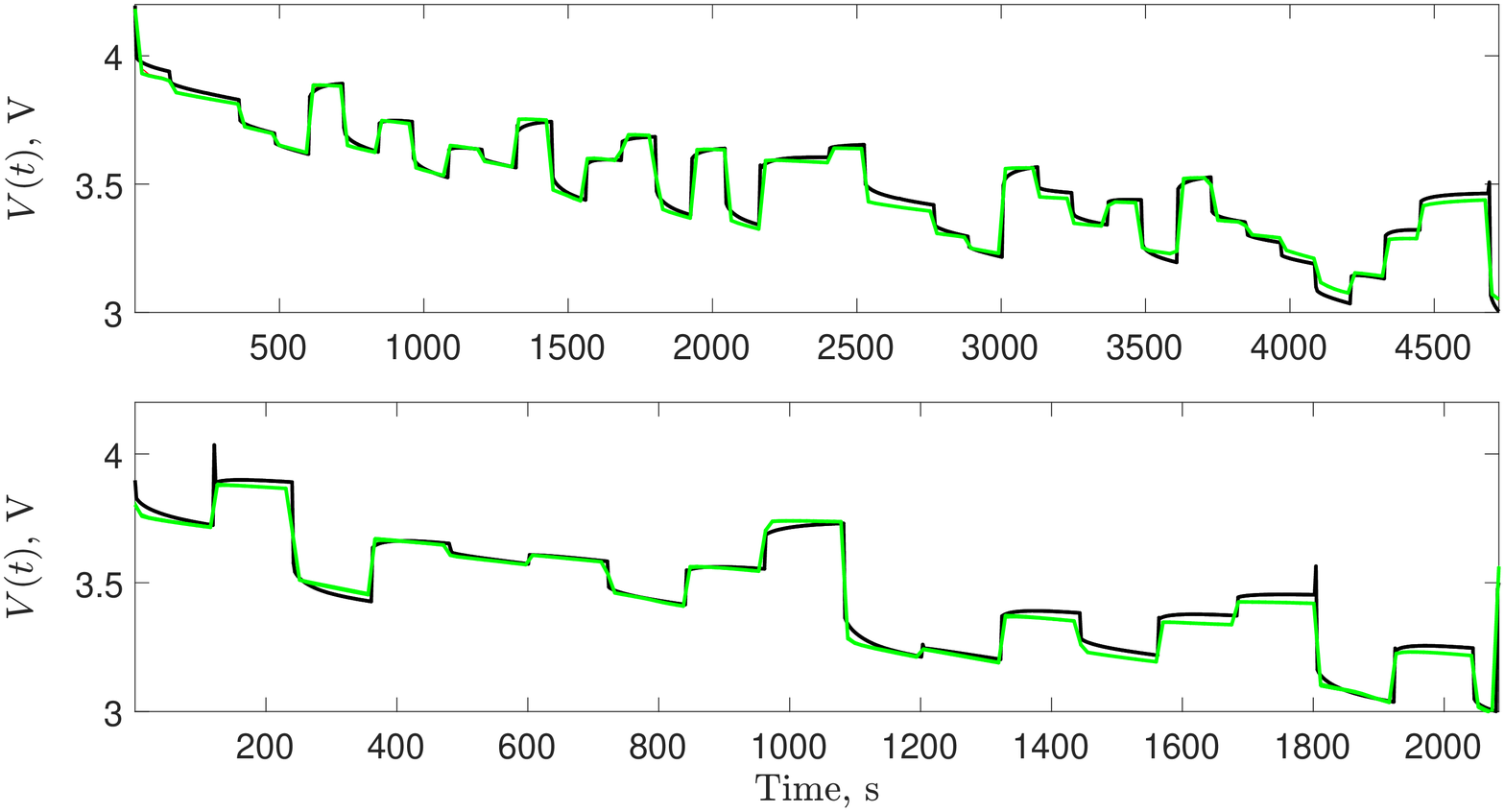} %the 0 in the beginning of the name should be eliminated 
\caption{Variable current discharge data (black) vs model predictions  (green). The model is best fit to the training constant-discharge data as in Figure \ref{fig:dataVsFit}.
Ten fit samples <1.02 min error (over training data) (green);  
The (time-)averaged error in prediction is $20mV$ i.e., about $2\%$ of the total voltage drop over the time of discharge. 
Computation time per discharge: $\sim 30s$.}
\label{fig:pulse1ans2}
\end{center}
\end{figure}
\begin{figure}[htp]
\vspace{-0.2cm}
\begin{center}
\includegraphics[width=3.3 in]{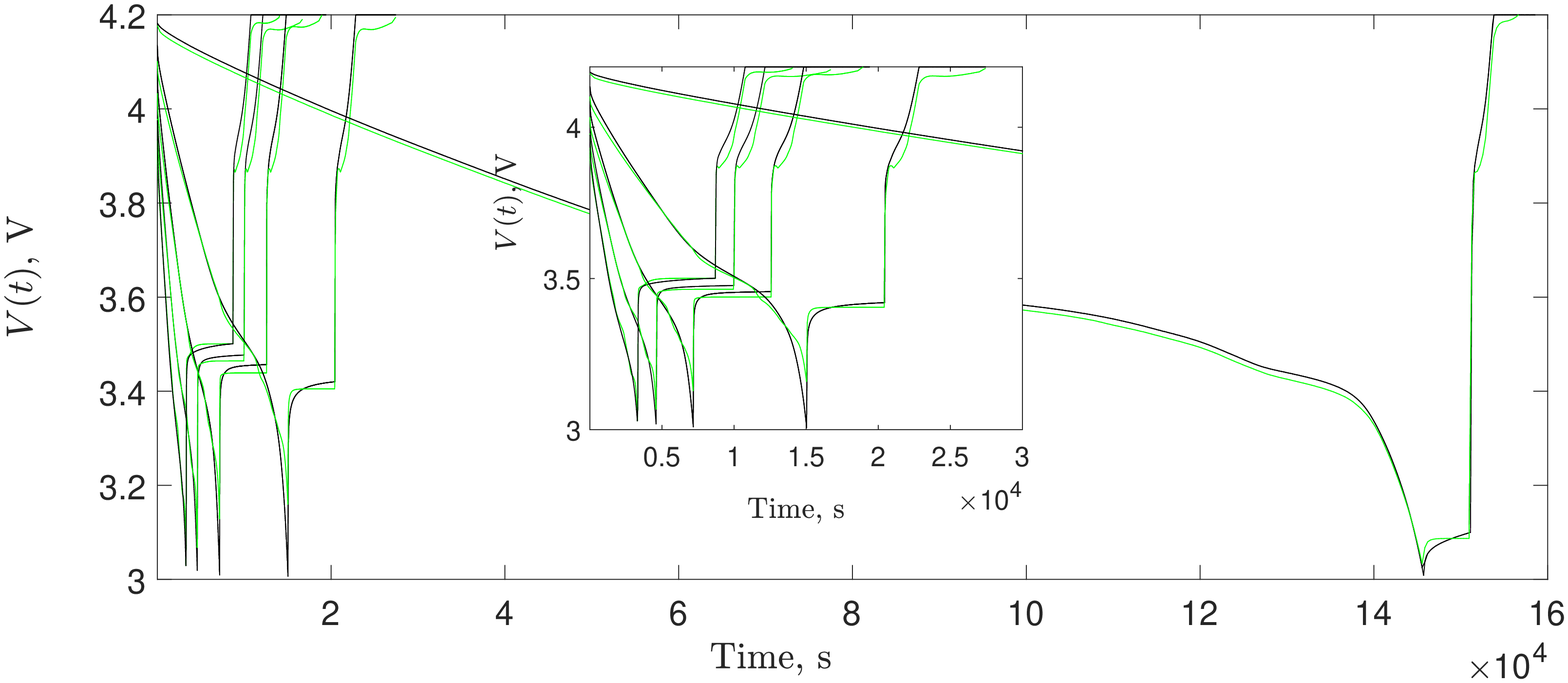} %the 0 in the beginning of the name should be eliminated 
\caption{Variable current discharging and charging data (black) vs model predictions  (green). The model is best fit to the training constant-discharge data as in Figure \ref{fig:chargingBFM20h8_2}.
A single sample from $\epsilon \le 1.02 \epsilon_{min|tr}$ ensemble (over training data) (green);  
The (time-)averaged error in prediction is $20mV$ i.e., about $2\%$ of the total voltage drop over the time of discharge. 
Computation time per discharge: $\sim 60s$.}
\label{fig:chargingBFM20h8_2}
\end{center}
\end{figure}

Each initial guess results in a generally different best-fit point in the parametric space, where the Nelder-Mead algorithm has stopped.  Therefore, an ensemble off initial guesses will lead to an ensemble of best-fit points. Upper row in Figure \ref{fig:BFM20h8b_presentation2} displays various projections of the resulting cloud of values in the parametric space (blue dots). We limit the plot to projections in the parametric subspace spanned by 6 parameters out of 9: the total Ohmic resistance $r$, the capacity $N_{Li,a}$, the  diffusion times $\tau_i$,  and specific exchange currents  $\tilde{I}_i$ for both the electrodes $i=a,c$. This particular choice is not arbitrary and will be discussed below. Different members of the best-fit ensemble correspond to slightly different RMSE. The member of the  ensemble  associated with the smallest RMSE $\epsilon_{min|tr}$ (conditioned on the training data set and ensemble of initial guesses) in the ensemble is plotted as the red dot. The cloud plotted in Figure \ref{fig:BFM20h8b_presentation2} corresponds to points in the 9-D parametric space associated with errors $\epsilon \le 1.02 \epsilon_{min|tr}$. For brevity we will omit using the word "cloud" in what follows and call "ensemble" both the ensemble and it's graphical representation. 

We note that the variations in the parametric values within the ensemble appear to be very large, and for some parameters larger by the factor of three than the associated RMSE $\epsilon_{min|tr}$ value. Figure \ref{fig:dataVsFit} shows the results of the fitting and constant-discharge testing of the model, fitted as explained above. The shown model predictions are made based on a random sample of $\sim 20$ parametric values drawn from the RMSE $\epsilon <1.02 \epsilon_{min|tr}$ ensemble. It is found that each model in the ensemble gives an averaged (over all the data) error of $\sim 2.2mV$, which corresponds  to $2.2\%$ of the total voltage drop. Prediction of the time of end of discharge is made with the average certainty of $1.15\%$. Remarkably, the variations between the models predictions are negligible. Figures \ref{fig:pulse1ans2} and \ref{fig:chargingBFM20h8_2} show the results of testing of the best-fit model against the variable current discharge data and charging/discharging data, respectively.  Each member of the  best-fit ensemble is seen to fit the variable-discharge and charging data to within a few percents as well.

It is clear that some aspects of the ensemble presented in the upper row of Fig. \ref{fig:BFM20h8b_presentation2}  are specific to the minimization algorithm. However, one  property appears to be characteristic of the RMSE landscape: even though the $\epsilon<1.02 \epsilon_{min|tr}$ ensemble of models show negligible variance in prediction of the constant and variable  discharges of interest, the variation of the associated values of parameters is very significant - of the order of the values themselves. This implies that the RMSE landscape is very broad in certain direction which makes parameters inference from the data an ill-posed problem. To quantify the observed property of the ensemble we adopted a slightly different RMSE minimization approach, where the Nelder-Mead minimization algorithm is forced to stop once  $\epsilon <1.02 \epsilon_{min|tr}$ is satisfied (stopping criterion). The resulting ensemble of parametric values is presented in the lower row of Fig. \ref{fig:BFM20h8b_presentation2}.

To characterize the RMSE landscape we performed the Principle Components Analysis (PCA) of the reduced covariance matrix associated with the ensemble $\epsilon <1.02 \epsilon_{min|tr}$.  We build the reduced covariance matrix as follows. The covariance matrix is defined by
\begin{eqnarray}
&&{Cov} _{  \bar{p}_{i}  \bar{p}_{j}}=\operatorname {E} [(  \bar{p}_{i}-\operatorname {E} [  \bar{p}_{i}])(  \bar{p}_{j}-\operatorname {E} [  \bar{p}_{j}])],
\end{eqnarray}
where $\mathbf {\bar{p}} =(  \bar{p}_{1},  \bar{p}_{2},...,  \bar{p}_{n})$ is a member of the best fit ensemble of nondimentionalized (by their characteristic scales  \footnote{Since the variances in the stiff and sloppy directions differ by orders of magnitude, the precise value of the characteristic scale is immaterial.}) parametric values $  \bar{p}_i$ in the $n$-D parametric space. The principle axes (PAs) of ${Cov} _{  \bar{p}_{i}  \bar{p}_{j}}$ are eigenvectors of the covariance matrix. The corresponding  eigenvalues are variances of the parameters in the directions of the PAs. For the calculation of the covariance matrix, as well as for the plot of the ensemble in the parametric space, Figure \ref{fig:BFM20h8b_presentation2}, we restricted ourselves to the subspace of parameters, spanned by 6 parameters out of 9 total: the total Ohmic resistance $r$, the capacity $N_{Li,a}$, the  diffusion times $\tau_i$,  and specific exchange currents  $\tilde{I}_i$ for both the electrodes $i=a,c$. The reason is that the values of left-out parameters $\Theta_{i,0}$, $i=a,c$, and $N_{Li,c}$, do not correlate with the values of the rest of the parameters for the following reasons: the value of the initial (fully charged battery) intercalation fraction in the anode $\Theta_{a,0}$ is determined by the given anode OCP;  the initial intercalation fraction $\Theta_{c,0}$ and the number of vacancies for intercalation $N_{Li,c}$ in the cathode were found by the preliminary sensitivity analysis to have essentially not effect on the fitting error. As a consequence, the covariance matrix in the full 9-D parametric space decomposes in two diagonal blocks: a 3-D block diagonal in the basis  $\Theta_{a,0}$ , $\Theta_{c,0}$ and $N_{Li,c}$ and the 6-D block associated with the subspace spanned by the rest of parameters, which was the main focus of the PCA.
\begin{figure}[htp]
\vspace{-0.2cm}
\begin{center}
\includegraphics[width=3.3 in]{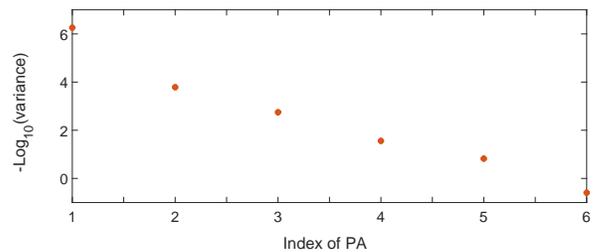} %the 0 in the beginning of the name should be eliminated 
\caption{Variances of the (nondimentionalized) ensemble, corresponding to RMSE $\epsilon \le 1.02 \epsilon_{min|tr}$ $\epsilon \le 1.02 \epsilon_{min|tr}$ (Figure \ref{fig:BFM20h8b_presentation2}), along the PAs. Variances are seen to vary by orders of magnitude. The corresponding PAs are given in Table \ref{table:PAs_F1_vs_F3b} (case F1). }
\label{fig:BFM20h8bVs20h8_spectrum}
\end{center}
\end{figure}
\begin{table*}[tbp]
\centering
\caption{Principle axes $\textbf{V}_i$ and the corresponding variances of the  $\epsilon \le 1.02 \epsilon_{min|tr}$ ensembles in the basis of nondimentionalized  physical parameters. Ordering: from smallest ($\textbf{V}_1$) to largest ($\textbf{V}_6$)  variance. }
\label{table:PAs_F1_vs_F3b}
\begin{tabular}{lllllllll}
\hline\hline
 Cases&\vline \ \ \  \ $\textbf{V}_i$    &\vline \ \ \   \  $\tau_c$ &\vline \ \ \  \  $\tilde{I}_a$   &\vline \ \ \  \  $\tilde{I}_c$  &\vline \ \ \  \  $r$   &\vline \ \ \  \ $N_{Li,a}$  &\vline \ \ \  \  $\tau_a$ &\vline \ \ \  \  $-\log_{10}(variance)$  \\ \hline \hline    
 &\vline \ \ \  \ $\textbf{V}_1$    &\vline \ \ \   -0.00&\vline \ \ \  \   0.001    &\vline \ \ \  \ 0.001     &\vline \ \ \   -0.004&\vline \ \ \  \ \textbf{-0.999}&\vline \ \ \  \ 0.040 &\vline \ \ \  \  6.26 \\ \hline
  &\vline \ \ \  \ $\textbf{V}_2$   &\vline \ \ \   -0.001&\vline \ \ \  \  -0.011   &\vline \ \ \   -0.009    &\vline \ \ \  \ 0.049 &\vline \ \ \  \ 0.04  &\vline \ \ \  \ \textbf{0.998} &\vline \ \ \  \  3.78 \\ \hline 
 F1&\vline \ \ \  \ $\textbf{V}_3$  &\vline \ \ \   \   0.031&\vline \ \ \  \ \textbf{-0.343}    &\vline \ \ \   \textbf{-0.296}    &\vline \ \ \  \ \textbf{0.89}  &\vline \ \ \   -0.006&\vline \ \ \   -0.05 &\vline \ \ \  \  2.75 \\ \hline
 &\vline \ \ \  \ $\textbf{V}_4$    &\vline \ \ \   -0.026&\vline \ \ \  \ \textbf{-0.498}    &\vline \ \ \    \textbf{-0.747}   &\vline \ \ \   \textbf{-0.439}&\vline \ \ \  \ 0.001 &\vline \ \ \  \ 0.01 &\vline \ \ \  \  1.56 \\ \hline 
 &\vline \ \ \  \ $\textbf{V}_5$    &\vline \ \ \   -0.015&\vline \ \ \  \ \textbf{-0.796}    &\vline \ \ \  \ \textbf{0.595}     &\vline \ \ \  \ \textbf{-0.109}&\vline \ \ \  \ 0.000 &\vline \ \ \  \ 0.002 &\vline \ \ \  \  0.82 \\ \hline 
 &\vline \ \ \  \ $\textbf{V}_6$    &\vline \ \ \   \  \textbf{0.999} &\vline \ \ \  \  -0.015  &\vline \ \ \    -0.001    &\vline \ \ \   -0.040&\vline \ \ \  \ 0.000 &\vline \ \ \  \ 0.003 &\vline \ \ \  \  -0.59 \\ 
\hline \hline                                               
\end{tabular}
\end{table*}

Figure \ref{fig:BFM20h8bVs20h8_spectrum} and Table \ref{table:PAs_F1_vs_F3b} present the variances and PAs, respectively, of the ensemble of fitted parametric values corresponding to RMSE $\epsilon <1.02 \epsilon_{min|tr}$.
Two remarkable observations can be made. First, we see that the variances of the ensemble in different directions of the nondimensionalized parametric space vary by many orders of magnitude. This means the while some combinations of the SPM parameters can be determined with a reasonable certainty, other combinations  are in practice left indeterminate by  fitting to the data. Second, the PAs of the ensemble do not all coincide with the axes, corresponding to the original parameters of the  model. Therefore, the combinations of parameters which can be determined with  relative certainty are not all interpretable in terms of the original parameters. To put the foregoing observation in a more general context, and to draw some practical conclusions, we review the theory of Sloppy-Models in the next section, as it appears that the SPM is a sloppy model in the sense of this theory.

\section{Sloppy models} \label{sec:sloppy}
The concept of Sloppy Models has been introduced in Ref.\cite{Brown2003} and subsequently developed into an elaborate theory with applications to  complex systems ranging from systems biology to particle accelerators \cite{Waterfall2006, Gutenkunst2007, Transtrum2010, Transtrum2011, Machta2013, OLeary2015, Niksic2016, Raman2017, Bohner2017, Raju2018, Mattingly2018}. Sloppy model of a complex system depend on a large number of "microscopic"  parameters of the system, but predict its "macroscopic" or "emergent" behavior. The term \textit{emergent behavior} is applied to an observed dynamics of a complex (multi-component) system, which is an outcome of the interaction of the system's parts and cannot be reduced to effects of noninteracting components. An emergent behavior depends smoothly on the control parameters (e.g., time) and few data points constrain the intermediate states variables \cite{Transtrum2010}. Sloppy models can be defined as models of a system's emergent behavior constrained by fewer effective data points than the number of parameters of the model. In this situation the problem of parameters inference becomes ill-posed. 

In practice, attempting to fit parameters of a sloppy model to the data, one encounters huge uncertainties of the best-fit values in certain ("sloppy" directions) of the parametric space and small variations in other ("stiff") directions. For the parameters nondimensionalized by their characteristic values the ratios of the variances span many orders of magnitude as observed in the preceding section for the SPM.

Formally, we can consider the sum of residuals cost function for the least squares fitting $C(\textbf{x},\bar{\textbf{p}})$:
\begin{eqnarray}
&&C=\sum_{i=1}^M \left[x_i-f_i(\bar{\textbf{p}})\right],  \label{eq:cost}
\end{eqnarray}
 where $x_i$, $i=1,2,...,M$ are the data points, $f_i$ vector of the model predictions and $\bar{p}_j$, $j=1,2,...,N$ are the values of parameters nondimensionlized by their characteristic scales. The minimal value of the cost is formally obtained at the best fit parametric  values $\bar{\textbf{p}}^0$ defined by 
\begin{eqnarray}
&&\frac{\partial C(\textbf{x},\bar{\textbf{p}})}{\partial \bar{p}_j}|_{\bar{\textbf{p}}^0}=0,  \quad H_{j,k}\equiv \frac{\partial^2 C(\textbf{x},\bar{\textbf{p}})}{\partial \bar{p}_j \partial \bar{p}_k}|_{\bar{\textbf{p}}^0}>0. \label{eq:minerror}
\end{eqnarray}
For Sloppy Models \cite{Waterfall2006} the Hessian is ill-conditioned; sloppy directions correspond to eigenvectors associated with vanishingly small eigenvalues. The spectrum of the Hessian is roughly evenly spaced on the logarithmic scale.

Importantly, due to the described properties of the error landscape associated with the sloppy models, small perturbations of the data, of the algorithm of the parameters inference and of the model itself  will generically lead to large variations of the best-fit values in the sloppy directions but will have small effect on the values of stiff parameters, Appendix \ref{app:linear}. As a consequence, inference of the values of sloppy parameters is not only infeasible but also meaningless.

Hessian characterizes the property of the cost function in the vicinity of its minimum in the parametric space. To characterize the global properties of a sloppy model Ref. \cite{Transtrum2010} introduced a geometrical object termed model manifold. The model manifold is defined as the locus of model predictions in the data space for all allowable parametric values. The data space is a metric space, with the distance between the two points $\mathbf{x}^j$ and $\mathbf{x}^k$ defined by the square root of the sum of square differences (Euclidean metric): 
\begin{eqnarray}
&&d_{jk}=\sqrt{\sum_i^M \left(x^j_i-x^k_i\right)^2},  \label{eq:dist}
\end{eqnarray}
where the number of data points $M$ determines the dimension of the data space. Comparing Eq.(\ref{eq:dist}) with Eq.(\ref{eq:cost}) one can see that the best fit value of the parameters corresponds to a point on the model manifold, which is closest to the data in the metric (\ref{eq:dist}). If the number of effective data points $M^{\prime}$ is much smaller than the number of model parameters $M^{\prime} \ll N$, as for sloppy models, the geometry of the model manifold is found to be that of hyperribons, displaying hierarchy of widths, where  the widths of the manifold in certain directions are much smaller than in others. Narrower widths correspond to sloppy directions in the parametric space and the edges of the hyperribon correspond to limiting values of the sloppy parameters, e.g., $0$ or $\infty$. This geometrical pictures suggests that the model manifold can be approximated by the edges of the hyperribons \cite{Transtrum2014}, leading to a systematic reduction of the sloppy model by taking the corresponding limits of the model's parameters in the sloppy directions. A reduced model depends on fewer sloppy parameters than the original one. Eventually, the reduction results in a model where where all the remaining parameters are practically stiff, i.e., can be inferred from fitting the reduced model to the data with fair amount of certainty. 

The method was termed Model Reduction by Manifold Boundaries and has been applied to date to a number of complex systems ranging from system biology \cite{Transtrum2014} and biophysics \cite{Bohner2017} to nuclear physics \cite{Niksic2016}. Although generally the application of the method involves numerically solving certain equations of motion on the model manifold, the limiting values of the parameters underlying the reduction can often be guessed based on the geometrical properties of the cost function (\ref{eq:cost}) in the vicinity of its minimum \cite{Transtrum2014}. This latter approach was adopted in the present work to derive a hierarchy of reduced models from the SPM. 

We found it convenient for visualization purposes to introduce a geometrical object in the parametric space which we termed the  best-fit manifold (BFM). An operational definition of the BFM is the locus of the parametric values which is consistent (to a prescribed accuracy in the metric (\ref{eq:cost})) with the data. In the sense of this definition, the ensemble of the best-fit values presented in Figure \ref{fig:BFM20h8b_presentation2} can be considered a sampling  of the BFM. Geometrical features of the BFM presented in the next section will guide our choice of the limiting values of parameters of the SPM for the model reduction. While some parameters are eliminated by taking the limits, values of other parameters are found by fitting the resulting reduced model to the data. Geometrically, taking all these limits can be seen as moving on the BFM in certain  directions of the parametric space. Staying on the BFM ensures that the fitting error of the reduced model is small, by the definition of the BFM, and elimination of sloppy parameters ensures that the reduced model depends on only stiff parameters.

\section{Reduction of the SPM}
\subsection{Best Fit Manifold of the SPM} \label{sec:nonlinear}
The Best-Fit Manifold (BFM) was defined in the preceding section as the locus of the parametric values which is consistent with the data in a certain sense. A natural way to quantify this consistency is in terms of the maximal value of the cost function (\ref{eq:cost}) or equivalently the RMSE of the model prediction over the BFM for the given data. In this case the BFM is just the union of the level sets of the cost function corresponding to its values smaller than or equal to this maximal value.
Ensemble of parametric values in Figure \ref{fig:BFM20h8b_presentation2} is a sampling of the BFM corresponding to the RMSE $\epsilon \le 1.02 \epsilon_{min|tr}$, where $\epsilon_{min|tr}$ is the minimal RMSE obtained in a series of minimization runs for varying initial guesses for the parameters, as explained in Section \ref{sec:fitting}. For such small deviation from the minimal value it can be expected that the linear analysis of Section \ref{sec:fitting}, including the PCA of the ensemble gives a faithful geometrical characterization of the BFM. Increasing the allowed deviation of the cost function from its local minimum one can build a family of nested BFMs. The linear analysis in Section \ref{sec:fitting} showed exponentially large scales separation for the BFM between the sloppy and stiff directions in the parametric space. This scales separation is expected to be preserved for BFM corresponding to larger allowed deviation of the cost function from its local minimum. Indeed, for sloppy models the number of effective data points $M^{\prime}$, corresponding to the number of stiff directions locally,  is smaller than the number of parameters $N$. In this case, the zero-level set of the cost function (\ref{eq:cost}) 
\begin{eqnarray}
&&C^{\prime}=\sum_{i=1}^{M^{\prime}} \left[x_i-f_i(\bar{\textbf{p}})\right]=0,  \label{eq:cost2}
\end{eqnarray}
comprises  a $(N-M^{\prime})$-dimensional manifold in the parametric space. By the definition of  effective data points, these level sets, for various choices of the effective data points, are local approximations of the BFMs corresponding to sufficiently small values of the cost function. It follows that the BFMs are quasi-$(N-M^{\prime})$-dimensional manifolds, approximately determined by the values of $M^{\prime}$ effective stiff parameters, corresponding locally to the $M^{\prime}$ stiff directions in the parametric space for the cost function (\ref{eq:cost}).

In Section \ref{sec:fitting} we estimated the local shape of the cost function (\ref{eq:cost}) level sets by the PCA of the best-fit ensembles corresponding to the RMSE $\epsilon \le 1.02 \epsilon_{min|tr}$. The variances were found to be exponentially larger in stiff directions compared to sloppy directions, Figure \ref{fig:BFM20h8bVs20h8_spectrum}. In this situation, generic for sloppy models (\cite{Waterfall2006}), the distinction between the sloppy and stiff parametric direction is, in practice, sharp. Analysis of effects of small perturbations of either the fitting procedure or the data on the best-fit ensemble in Appendix \ref{app:linear} has shown that it shifts significantly in the sloppy directions in parametric space, while  changes in the stiff directions are negligible. This picture suggests a operational definition of {the} stiff directions  as \textit{directions which are stiff under typical perturbations of the experimental conditions and the fitting procedure, which negligibly affects the fitting quality}. Based on this definition, we assessed the number of stiff parameters for the SPM: $M^{\prime}=3$. Two out of the three parameters were found to be the original ("microscopic") parameters of the SPM: the diffusion time $\tau_a$ and the number of Li ions $N_{Li,a}$ in anode. Remarkably, the remaining stiff parameter $\tilde{r}$ was found to be a combination of the total Ohmic resistance $r$  and the electrodes specific exchange currents $\tilde{I}_i$, $i=a,c$. In the light of the preceding discussion, the BFMs for the 9-parametric SPM with the value of parameter $\Theta_{a,0}$ fixed by the anode OCP are expected to be quasi-5-dimensional manifolds, defined approximately by values of 3 stiff parameters and $\Theta_{a,0}$. 
\subsection{Reduction of the SPM} \label{sec:reduction}
\begin{figure}[htp]
\vspace{-0.2cm}
\begin{center}
\includegraphics[width=3.4 in]{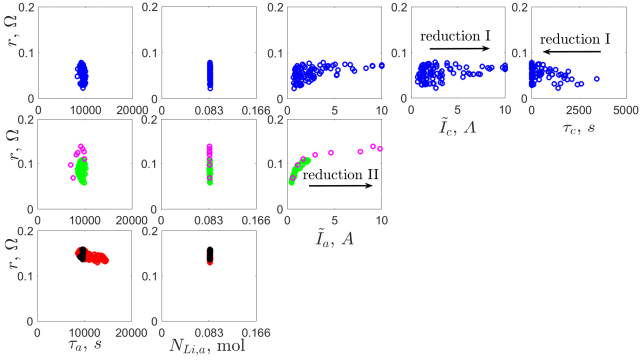} %the 0 in the beginning of the name should be eliminated pulse1ans2
\caption{Ensembles corresponding to RMSE $\approx 1.1 \epsilon_{min|tr}$ for full SPM (upper row, blue), RMI (middle row, green) and RMII (bottom row, red). The number of sloppy parameters in the RMI is reduced to just one, corresponding to the effective dimension 1 of the resulting BFM. The values of the stiff parameters of the full SPM are preserved. A random sample of the $\approx 1.3 \epsilon_{min|tr}$ ensemble for RMI (magenta) is shown to guide the path in the parametric space ($\tilde{I}_a \rightarrow \infty$) for the subsequent reduction of the model. The number of sloppy parameters in the RMII is reduced to zero, corresponding to the effective dimension zero of the BFM. The apparent increase of uncertainty of the $\tau_a$ is fortuitous and explained by the figure \ref{fig:zebra}. Black dots correspond to the RMSE $\approx 1.1 \epsilon_{min|tr}$ ensemble, originating from the initial guesses for $\tau_a$ and $N_{i,a}$ in the vicinity of the best fit obtained in the full SPM and RMI. The corresponding values of $\tau_a$ and $N_{i,a}$ are preserved across the hierarchy of the reduced modes as expected for the stiff parameters.}
\label{fig:full_vs_red2_3}
\end{center}
\end{figure}

Viewing the BFM corresponding to a larger allowed deviation of the cost function from its local minimum gives ideas for possible limiting values of the SPM parameters, for which a simpler reduced model can be obtained, while staying on the BFM. Staying on the BFM  ensures that the resulting model's prediction fit the data reasonably well. The upper row in Figure \ref{fig:full_vs_red2_3} shows projections of the ensemble of best-fit values, corresponding to the TMSE $\epsilon \le 1.1 \epsilon_{min|tr}$, where $\epsilon_{min|tr}$ is the minimal error. The values of the specific exchange currents $\tilde{I}_i$, $i=a,c$ are seen to reach $10A$ on the BFM. The cycling data used for fitting the model parameters, Figure \ref{fig:dataVsFit}, corresponds to constant discharge currents $I \lesssim 2A$. Therefore, the values of $10A$ for the exchange currents observed on the BFM, are significantly larger than the operational currents and suggest that taking limits $\tilde{I}_i \rightarrow \infty$ may be appropriate for model reduction, provided they exist and are nonsingular. Similarly, the values of the cathode diffusion time $\tau_c$ are seen to reach values on the BFM which are much shorter than the characteristic discharge time in the cycling data. This suggests that $\tau_c \rightarrow \infty$ may also be an appropriate limit for the model reduction, provided it exists and is nonsingular.

\begin{figure}[htp]
\vspace{-0.2cm}
\begin{center}
\includegraphics[width=3.3 in]{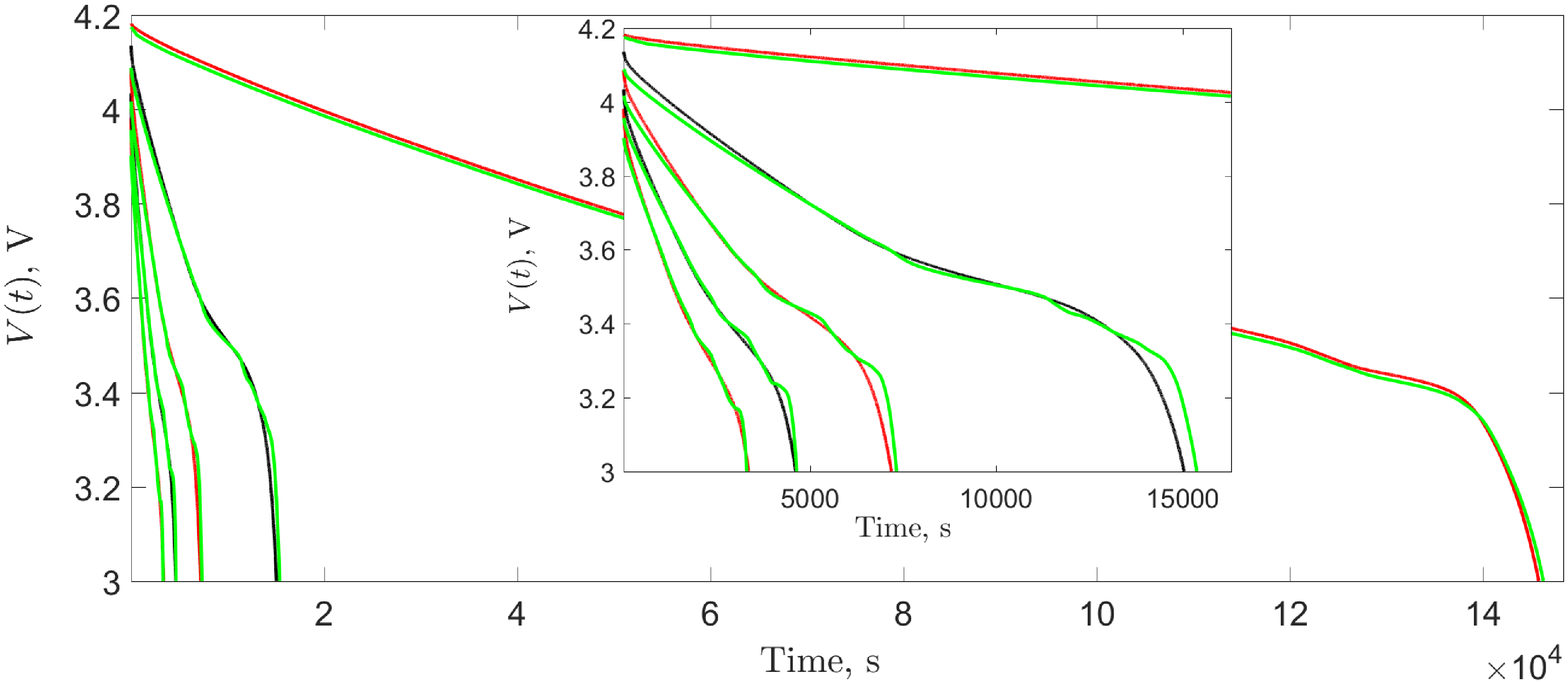} %the 0 in the beginning of the name should be eliminated pulse1ans2
\caption{Training (red) and testing (black) data vs RMI (green) for discharge currents: $2.0,1.5,1.0,0.5,0.055A$. 
The training data set is generated by $2.0,1.0,0.055A$ discharge currents. $20$ model predictions (trajectories) are made based on $20$ random sets of parametric values within the range RMSE $\epsilon<1.02 \epsilon_{min|tr}$ - the minimal RMSE obtained for the training data; $\epsilon_{min|tr} \approx 2.5mV$. RMSE averaged over all the data is $2.5\%$ for each parametric set (the variation between the sets are negligible); average (over the data) error in the time of the end of discharge $t_{EOD}$ prediction is $1.3\%$(the variation between the sets are negligible). Computation time per trajectory: $\sim 0.05-0.1s$.}
\label{fig:red2}
\end{center}
\end{figure}

The corresponding model reduction was performed in two steps. 
The Reduced Model I (RMI) is obtained from Eqs.(\ref{eq:governing}) as the limiting model for $\tilde{I}_c \rightarrow \infty$ and $\tau_c \rightarrow 0$:
\begin{eqnarray} 
&&V(t)=\Delta \overline{\phi^{eq}_c}(t)-\Delta \phi^{eq}_a(\bar{\theta}_a  )-I_a \left(r_a+r_c\right)\nonumber \\
&&-\frac{2 k_B T}{e} \ln\left({\chi_a  +\sqrt{(\chi_a  )^2+1}}\right), \label{eq:theta_c_1}\\
&&\Delta \overline{\phi^{eq}_c}(t) \equiv U_{OCP}(y(t))+\Delta \phi^{eq}_a(y(t)), \label{eq:theta_c_2} \\
&& y(t)=1-\int_0^t \frac{I_a(t') dt'}{e N_{Li,a}}, \label{eq:theta_c_3} \\
&&  \chi_a  (\bar{\theta}_a)=\frac{I_a}{\tilde{I}_a \left({\theta}_a  \right)^{ \frac{1}{2}} \left(1-{\theta}_a  \right)^{ \frac{1}{2}}}, \nonumber \\
&& I_a > 0 \ \ for \ \ discharge, \quad {\theta}_a=\bar{\theta}_a \Theta_{a,0},  \nonumber \\
&&Parameters \  (3):\  r=r_a+r_c, \ \tilde{I}_a, \ \Theta_{a,0} , \nonumber
\end{eqnarray}
where expressions (\ref{eq:theta_c_1})-(\ref{eq:theta_c_3}) follow from Eq.(\ref{eq:solbSPM2_sum}) and (\ref{eq:OCV_thetac}) in the limit of $\tau_c \rightarrow 0$ and 
\begin{eqnarray} 
&&\bar{\theta}_a(t)=1-{\mathcal{L}}^{-1}\left[\frac{\text{sign}(I_a){\mathcal{L}}\left[\bar{\zeta}_a( t')\right] }{  \sqrt{s} \coth {\sqrt{s}}-1}\right] \left(\frac{t}{\tau_a}\right), \\
&&\bar{\zeta}_a(t/\tau_a)= \frac{I(t) \tau_a }{3 e N_{Li,a}},   \\
&&Parameters \ (2):\  \tau_a, \ N_{Li,a}; \nonumber 
\end{eqnarray}
In this model, $4$ of the original $9$ parameters have been eliminated. In the remaining $5$ parameters $3$ are stiff and one is sloppy, and $\Theta_{a,0}$ is kept fixed by the anode OCP. The minimal RMSE of the model is slightly larger than in the original SPM. The results of the fitting is provided in Figure \ref{fig:red2}. Comparison of the errors in prediction of voltage and time of end of discharge is given in table \ref{table:Q_of_Fits_red2}. It's seen that the errors are just a bit larger than for the full model. However inspection of the figure \ref{fig:full_vs_red2_3} shows that the resulting BFM is $1$-D as expected. 
\begin{figure}[htp]
\vspace{-0.2cm}
\begin{center}
\includegraphics[width=3.3 in]{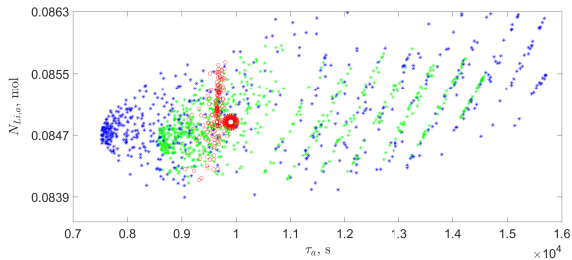} %the 0 in the beginning of the name should be eliminated pulse1ans2
\caption{RMII. Projection of the ensembles corresponding to RMSE $\approx 1.1 \epsilon_{min|tr}$ (green) and $\approx 1.2 \epsilon_{min|tr}$ (blue) on the $(\tau_a, N_{Li,a})$ plane. The large red circle corresponds to the minimal RMSE, and coincides essentially with the minimal value for the full model and RMI. The observed stripes correspond to a series of local minima of the RMSE, separated by "ridges" of high RMSE of at least $\approx 1.2 \epsilon_{min|tr}$. In contrast to the effective degeneracy of the minimum for the sloppy full SPM (quasi-$5$-D BFM) and reduced model I (quasi-$1$-D BFM) the separated minima cannot be reached by a small perturbation of the data, model or fitting  procedure. Therefore, the physical minimum is isolated and the physically relevant width of the  ensemble for RMSE $\approx 1.1 \epsilon_{min|tr}$ is limited by first ridge separating the minimum from the next one. This width is found to be about $10\%$ as for the full model and RMI. To converge to the physical minimum the fitting procedure should start at the initial guesses for $\tau_a$ and $N_{i,a}$ in the vicinity of the best fit obtained in the full model and RMI. Small red circles correspond to the ensemble with RMSE $\approx 1.1 \epsilon_{min|tr}$, originating from such an initial guess.}
\label{fig:zebra}
\end{center}
\end{figure}
\begin{table*}[tbp]
\centering
\caption{Quality of fits. Case F1}
\label{table:Q_of_Fits_red2}
\begin{tabular}{lllllll}
\hline\hline
   &\vline \ \ &\vline \ \ 2A &\vline \ \  1.5A &\vline \ \  1A &\vline \ \  0.5A &\vline \ \  0.055A \\ \hline
Full Model  \ \  &\vline \ \ \begin{tabular}[c]{@{}l@{}}$\epsilon_{t_{EOD}}$\\   $\epsilon_{V(t)}$\end{tabular} &\vline \ \  \begin{tabular}[c]{@{}l@{}}-1.2 \%\\   1.5e-02 V\end{tabular} &\vline \ \  \begin{tabular}[c]{@{}l@{}}0.5 \%\\   2.1e-02 V\end{tabular} &\vline \ \  \begin{tabular}[c]{@{}l@{}}1.8 \%\\   2.9e-02 V\end{tabular} &\vline \ \  \begin{tabular}[c]{@{}l@{}}2 \%\\   3.3e-02 V\end{tabular} &\vline \ \  \begin{tabular}[c]{@{}l@{}}-0.1 \%\\   1.1e-02 V\end{tabular}\\ \hline
Reduced model I &\vline \ \ \begin{tabular}[c]{@{}l@{}}$\epsilon_{t_{EOD}}$\\   $\epsilon_{V(t)}$\end{tabular} &\vline \ \  \begin{tabular}[c]{@{}l@{}}-1.4 \%\\   1.4e-02 V\end{tabular} &\vline \ \  \begin{tabular}[c]{@{}l@{}}0.6 \%\\   2.5e-02 V\end{tabular} &\vline \ \  \begin{tabular}[c]{@{}l@{}}1.9 \%\\   3.6e-02 V\end{tabular} &\vline \ \  \begin{tabular}[c]{@{}l@{}}2.3 \%\\   3.7e-02 V\end{tabular} &\vline \ \  \begin{tabular}[c]{@{}l@{}}0.3 \%\\   1.1e-02 V\end{tabular} \\
\hline
Reduced model II &\vline \ \ \begin{tabular}[c]{@{}l@{}}$\epsilon_{t_{EOD}}$\\   $\epsilon_{V(t)}$\end{tabular} &\vline \ \  \begin{tabular}[c]{@{}l@{}}-0.6 \%\\   2.0e-02 V\end{tabular} &\vline \ \  \begin{tabular}[c]{@{}l@{}}0.8 \%\\   3.4e-02 V\end{tabular} &\vline \ \  \begin{tabular}[c]{@{}l@{}}2.3 \%\\   5.0e-02 V\end{tabular} &\vline \ \  \begin{tabular}[c]{@{}l@{}}2.6 \%\\   5.0e-02 V\end{tabular} &\vline \ \  \begin{tabular}[c]{@{}l@{}}0.2 \%\\   0.71e-02 V\end{tabular} \\
\hline\hline                                                          
\end{tabular}
\end{table*}

The next iteration of reduction results in reduced model II (RMII) and is obtained from Eqs.(\ref{eq:theta_c_1})-(\ref{eq:theta_c_3})  as the limiting model for $\tilde{I}_a \rightarrow \infty$ (see Figure \ref{fig:full_vs_red2_3} for the guideline for the reduction):
\begin{eqnarray} \label{eq:governing_red_2}
&&V(t)=\Delta \overline{\phi^{eq}_c}(t)-\Delta \phi^{eq}_a(\bar{\theta}_a  )-I_a \left(r_a+r_c\right),  \\
&&\Delta \overline{\phi^{eq}_c}(t) \equiv U_{OCP}(y(t))+\Delta \phi^{eq}_a(y(t)), \nonumber \\
&& y(t)=1-\int_0^t \frac{I_a(t') dt'}{e N_{Li,a}}, \nonumber \\
&&Parameters \  (1):\  r=r_a+r_c, \nonumber 
\end{eqnarray}
and 
\begin{eqnarray} 
&&\bar{\theta}_a(t)=1-{\mathcal{L}}^{-1}\left[\frac{\text{sign}(I_a){\mathcal{L}}\left[\bar{\zeta}_a( t')\right] }{  \sqrt{s} \coth {\sqrt{s}}-1}\right] \left(\frac{t}{\tau_a}\right), \\
&&\bar{\zeta}_a(t/\tau_a)= \frac{I(t) \tau_a }{3 e N_{Li,a}}. \nonumber \\
&&Parameters \ (2):\  \tau_a, \ N_{Li,a}. \nonumber
\end{eqnarray}
In this model, $6$ of the original $9$ parameters have been eliminated. The remaining $3$ parameters $\tau_a, N_{Li,a}$ and $\tilde{r}$ are stiff. A spurious increase of the uncertainty in $\tau_a$ is seen in Figure \ref{fig:full_vs_red2_3}, but not observed for either full SPM or RMI.  Projection of the ensembles corresponding to RMSE $\approx 1.1 \epsilon_{min|tr}$ (green)  on the $(\tau_a, N_{Li,a})$ plane, Figure \ref{fig:zebra}, shows a peculiar striped pattern in the ensemble. It apparently corresponds to sequence of local minima of the cost function separated by ridges of high values of the RMSE. Ensemble corresponding to $\approx 1.2 \epsilon_{min|tr}$ (blue) shows the same location of the ridges. Ensemble corresponding to $\approx 1.5 \epsilon_{min|tr}$ (not shown) does not longer present the striped pattern giving an upper bound for the ridges height. Since these  multiple  minima of the RMSE are separated by the high cost ridges, they cannot be reached by small perturbations of the data, model or fitting  procedure. In this sense the absolute minimum of the RMSE (the red circle in figure \ref{fig:zebra}) which coincides  with the minimal value for the full model and RMI, should be considered as the physically meaningful best fit value. 
Since $\tau_a$ is a stiff parameter in (both full and) RMI, its best fit value for the RMI can be used as an initial guess for fitting the RMII. This best fit can be obtained in just few iterations of the fitting procedure for the RMI and then used as the initial guess for the RMII. 

\begin{figure}[htp]
\vspace{-0.2cm}
\begin{center}
\includegraphics[width=3.3 in]{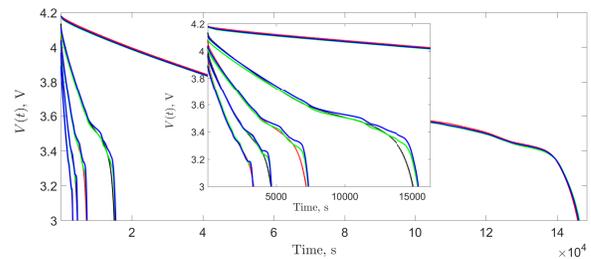} %the 0 in the beginning of the name should be eliminated pulse1ans2
\caption{Training (red) and testing (black) data vs RMII model (blue) for discharge currents: $2.0,1.5,1.0,0.5,0.055A$. 
The training data set is generated by $2.0,1.0,0.055A$ discharge currents. Ten models' predictions (trajectories) are made based on $10$ random sets of parametric values within the range RMSE $\epsilon<1.02 \epsilon_{min|tr}$, corresponding to the physical local minimum in Figure \ref{fig:zebra}; $\epsilon_{min|tr} \approx 3.2mV$. RMSE averaged over all the data is $3.2\%$ for each parametric set (the variation between the sets are negligible); average (over the data) error in the time of the end of discharge $t_{EOD}$ prediction is $1.5\%$(the variation between the sets are negligible). Computation time per trajectory: $\sim 0.05-0.1s$. The RMII fit (blue) is compared to RMI fit (green) reproduced from the Figure \ref{fig:red3}. }
\label{fig:red3}
\end{center}
\end{figure}

\begin{figure}[htp]
\vspace{-0.2cm}
\begin{center}
\includegraphics[width=3.3 in]{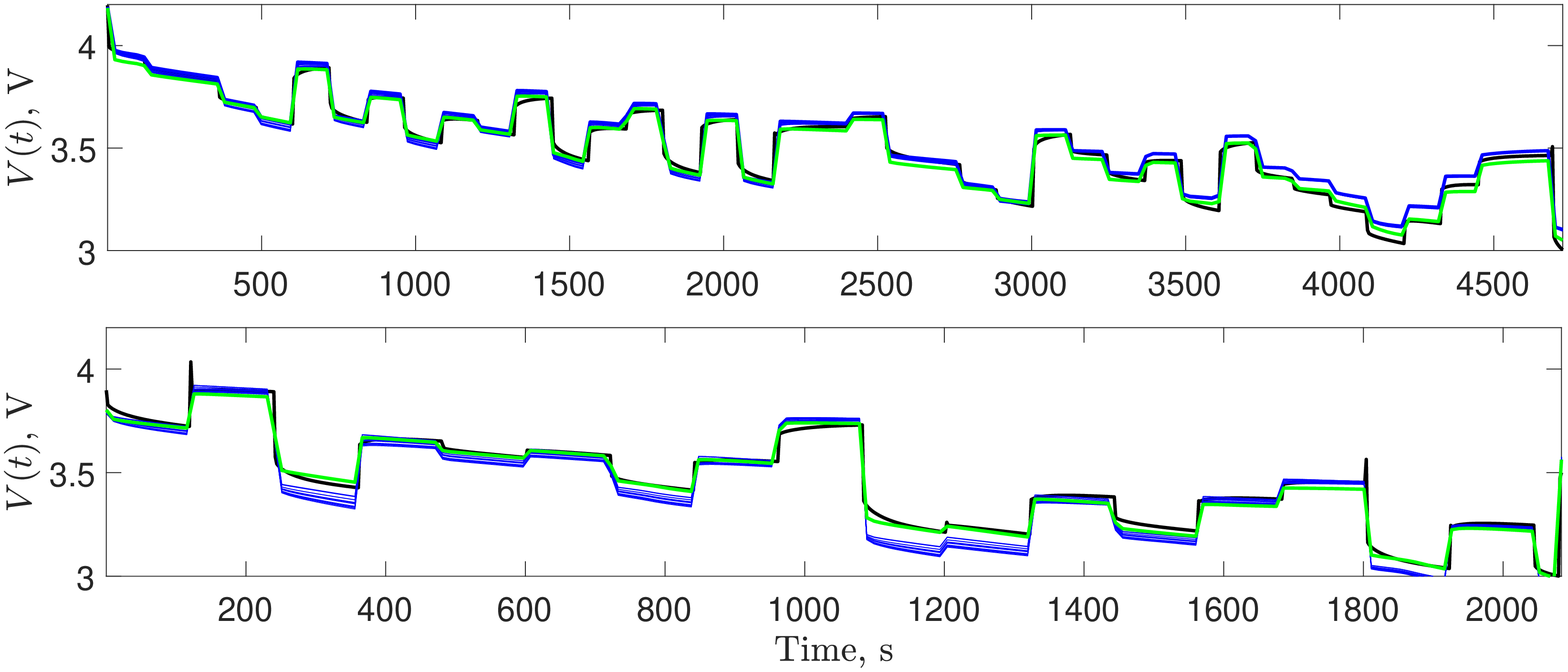} %the 0 in the beginning of the name should be eliminated 
\caption{Variable current discharge data (black) vs model predictions for RMII (blue) vs SPM. Both RMII and full SPM model are best-fit to the training constant-discharge data as in Figure \ref{fig:dataVsFit}.
Ten  samples with RMSE $\epsilon<1.02 \epsilon_{min|tr}$ are plotted (green);  
The time- and samples-averaged error in  RMII prediction is $35mV$ and $50mV$ for the pulses in the upper and lower panels, respectively,  corresponding to  $3.5\%$ and $5\%$, respectively, of the total voltage drop over the time of discharge. The corresponding predictions of the SPM (green) give $2\%$ average error.
Computation time per discharge for RMII: $\sim 30s$.}
\label{fig:pulse1ans2_paper2}
\end{center}
\end{figure}

The resulting  RMSE of the RMII is slightly larger than in the RMI. The results of the fitting is provided in Figures \ref{fig:red3} and \ref{fig:pulse1ans2_paper2}. Comparison of the errors in prediction of voltage and time of end of discharge is given in table \ref{table:Q_of_Fits_red2}. It's seen that the errors are just a bit larger than for the full SPM and the RMI. However inspection of the figure \ref{fig:full_vs_red2_3}  and \ref{fig:zebra} shows that the resulting BFM is effectively $0$-D as expected. 

Reduction of the SPM leads to a moderate decrease in its fidelity in predicting the cycling behavior of the battery. On the other hand, a reduced model depends on fewer unidentifiable parameters, therefore providing a better tool for characterizing the  battery's physical state. In a fully reduced model - RMII - this characterization is accomplished in terms of values of just three stiff parameters $\tau_a$, $N_{Li,a}$ and $r$, which are provided by fitting the model to the data. Their values can be inferred with a high degree of certainty using the RMII. It is crucially important to note, that although formally $r$ in the RMII stands for the Ohmic resistance, its inferred value does not represent the true value of Ohmic resistance of the battery, but rather an effective resistance $\tilde{r}$, which is a function of the original parameters $r,\tilde{I}_a, \tilde{I}_c$ and which value is obtained by moving on the BFM of the full SPM in the subspace spanned by $r,\tilde{I}_a, \tilde{I}_c$. In this respect the performed reduction of the SPM and similar reductions of sloppy models in general are very different from conventional simplifications of models based on neglecting small parameters. In the latter case the remaining parameters of the model retain their original physical meaning and their best-fit values can be considered as fair estimates of the actual parametric values. In the case of sloppy models, parameters are neglected not because their actual values are small, but because their effect on predicting the system behavior is negligible.  The  remaining parameters should no longer be interpreted as original ("microscopic") parameters of the sloppy model, but should be considered effective parameters, depending on a number of original parameters.

\section{Conclusions}
Best-Fit Manifold (BFM) of the Single-Particle Model (SPM) of Li-ion battery has been characterized numerically. BFM was defined as locus of the values of the SPM parameters consistent with the cycling data. Local analysis of the BFM revealed exponentially large ratios of the best-fit parameters variations in certain (sloppy) directions in the parametric space, compared to other (stiff) directions. Due to this exponential discrepancy, parametric values in the sloppy direction cannot be inferred with a meaningful certainty from fitting the SPM to the cycling data. This property is known to be shared by a class of so-called Sloppy Models. Theory of sloppy models provide a formal pathway for a sloppy model reduction to a model depending on just stiff parameters. In many cases, a pathway can be guessed based on geometrical insights. Such is the case of the SPM, where geometrical properties of the BFM  allowed us to develop a hierarchy of the reduced models starting at the full model. Reduced model I (RMI) depends on fewer sloppy parameters, therefore leading to a more concise battery characterization. Fully reduced model II (RMII) depends on only three effective stiff parameters, the values of which can be inferred with high certainty by fitting the RMII to cycling data. As a consequence, the RMII provides a tool for characterization of the battery physical state in terms of just three parameters. The effective parameters of the RMII are the battery's capacity $N_{Li,a}$, the anode diffusion time $\tau_a$  and an effective Ohmic resistance $\tilde{r}$. The first two parameters are original parameters of the SPM, which can be inferred with certainty using the full SPM, and retain the original meaning in the process of model reduction. The third parameter $\tilde{r}$ is an effective parameter and not an original parameter of the SPM. It is an unknown function of three original parameters of the SPM: the total Ohmic resistance $r$ and the specific exchange currents at anode and cathode, $\tilde{I}_i$, $i=a,c$. As a consequence, it seems that the state of the battery cannot be characterized in terms of the original parameters of the SPM, based on the cycling data, but characterization can be accomplished in terms of effective parameters. In particular, it appears that modeling aging and degradation of Li ion battery based on cycling data alone can only be done in terms of the effective parameters, i.e., phenomenologically. Such an analysis is currently work in progress which will be reported elsewhere. 

The present work focuses on a particular model of the Li ion cells - the Single Particle Model. However,  based on our understanding of the root cause for the SPM "sloppiness" with respect to cycling data, we expect that more elaborate and high fidelity models, depending on a larger number of parameters will display "sloppiness" to an even greater extent. Application of a similar analysis to these model will lead to their reduction and building a hierarchy of reduced models in a systematic manner. In this sense, we believe that results of the present work have general character and will be useful in modeling and parameter inference  for Li ion cells.

\appendix

\section{SPM equations} \label{sec:SPM_eqs}

\subsection{Solid phase diffusion}
Evolution of the ions intercalation fraction $\Theta_i(r,t)$ in $i$th electrode ($i=a,c$, for anode and cathode, respectively) particle, corresponding to the picture above is expressed as follows:
\begin{eqnarray}
&&\frac{\partial \Theta_i}{\partial t}=\frac{1}{r^2} \frac{\partial}{\partial r} \left(r^2 D_i \frac{\partial}{\partial r}\Theta_i\right), \label{eq:diffSPM} \\
&&D_i\frac{\partial\Theta_i}{\partial r}|_{r=R_i}=-\frac{I_i(t) }{4 \pi R_i^2 e K_i  m_i},   \\
&&\frac{\partial\Theta_i}{\partial r}|_{r=0}=0, 
\end{eqnarray}
where the current sign convention is $I_a=-I_c$, $\left|I_i\right|=I$ and $I_a>0$ for discharging current, $D_i$ is solid phase diffusivity of $Li$, $R_i$ is the particle radius, $K_i$ the number of particles and $m_i$ is concentration of intercalation sites in electrode $i=a,c$.

Having in mind applications where at $t=0$ the intercalation fraction of ions in both electrodes is uniform, we fix the initial conditions by $\Theta_i(r,0)=\Theta_{i,0}$, for $r$-independent initial intercalation fractions $\Theta_{i,0}$, $i=a,c$.  Transforming to dimensionless variables $\bar{r} = r/R_i$, $\bar{t} = t/\tau_i$ where the \textit{diffusion time} is defined by $\tau_i\equiv R_i^2/D_i$, we obtain solution of Eqs.(\ref{eq:diffSPM})  in form of inverse Laplace transform for $t>0$:
\begin{eqnarray}
&&\Theta_i(\bar{r},\bar{t})=\Theta_{i,0} \nonumber \\
&&-{\mathcal{L}}^{-1}\left[\frac{{\mathcal{L}}\left[\zeta_i(\bar{t}')\right] \sinh{\bar{r} \sqrt{s} }}{\bar{r}  \left(\sqrt{s} \cosh {\sqrt{s}}-\sinh \sqrt{s}\right)}\right](\bar{t}), \label{eq:InvLaplaceSPM} \\
&&\zeta_i(\bar{t})= \frac{I_i(\tau_i \bar{t}) \tau_i}{3eM_{Li,i}}, \quad  \quad M_{Li,i} \equiv \frac{4}{3} \pi R_i^3K_i  m_i. \nonumber
\end{eqnarray}
We are interested in the solution at the boundary $\bar{r}=1$, which effects the measured voltage through the electrochemical kinetics:
\begin{eqnarray}
&&\theta_i(\tau_i\bar{t})\equiv \Theta_i(1,\bar{t}) \nonumber \\
&&=\Theta_{i,0}-{\mathcal{L}}^{-1}\left[\frac{{\mathcal{L}}\left[\zeta_i(\bar{t}')\right] }{  \sqrt{s} \coth {\sqrt{s}}-1}\right](\bar{t}). \label{eq:solbSPM}
\end{eqnarray}
Defining a \textit{normalized intercalation fraction} at the boundary by:
\begin{eqnarray}
&&\bar{\theta}_a=\frac{{\theta}_a}{\Theta_{a,0}}, \quad \bar{\theta}_c=\frac{{1-\theta}_c}{1-\Theta_{c,0}}, \label{eq:theta_norm}
\end{eqnarray}
so that $\theta_i=0$($\theta_i=1$) correspond to fully discharged (charged) cell, and transforming to original time $t$, we obtain:
\begin{eqnarray}
&&\bar{\theta}_i(t)=1-{\mathcal{L}}^{-1}\left[\frac{\text{sign}(I_a){\mathcal{L}}\left[\bar{\zeta}_i( t')\right] }{  \sqrt{s} \coth {\sqrt{s}}-1}\right] \left(\frac{t}{\tau_i}\right), \label{eq:solbSPM2}\\
&&\bar{\zeta}_i(t/\tau_i)= \frac{I(t) \tau_i }{3 e N_{Li,i}}, \quad \tau_i\equiv\frac{R_i^2}{D_i} \nonumber \\
&&N_{Li,a}={M_{Li,a}\Theta_{a,0}}, \quad N_{Li,c}={M_{Li,c}\left(1-\Theta_{c,0}\right)}, \nonumber  \\
&&M_{Li,i} \equiv \frac{4}{3} \pi R_i^3K_i  m_i, \nonumber 
\end{eqnarray}
where $I_a(t)=-I_c(t)$, $I\equiv \left|I_i\right|$, and $I_a(t)>0$ for a discharging current. The physical meaning of $N_{Li,i}$ for $i=a,c$ is the total number of Li ions in the  anode and intercalation vacancies in cathode, respectively, in the fully charged cell. The physical meaning of $\bar{\zeta}_i$ is the fraction of the total charge of available ions in anode $e N_{Li,a}$ ($i=a$) or of available intercalation vacancies in cathode $e N_{Li,c}$ ($i=c$) transferred by the current $I$ over the diffusion time $\tau_i$.

If $t=0$ corresponds to the beginning of discharge, and the discharge proceeds at constant current $I$ for $0<t<t_{EOD}$, followed by the recovery stage  $I=0$, the solution can be expressed as follows:
\begin{eqnarray}
&&\theta_i(t)=\Theta_{i,0}\nonumber \\
&&-\zeta_i{\mathcal{L}}^{-1}\left[\frac{1-e^{-s t_{EOD}/\tau_i} }{ s \left(\sqrt{s} \coth {\sqrt{s}}-1\right)}\right] \left(\frac{t}{\tau_i}\right), \label{eq:gentEOD} \\
&&\zeta_i=\frac{I_i }{I_{i,eff}},  \quad \tau_i\equiv \frac{R_i^2}{D_i} \nonumber \\
&&\ I_{i,eff}\equiv \frac{3eM_{Li,i}}{\tau_i},  \quad M_{Li,i} \equiv \frac{4}{3} \pi R_i^3K_i  m_i. \nonumber
\end{eqnarray}
A computationally friendly form of Eq.(\ref{eq:gentEOD}) for $t<t_{EOD}$ is found to be:
\begin{eqnarray}
&&\theta_i(t)=\Theta_{i,0}\nonumber \\
&&-\zeta_i{\mathcal{L}}^{-1}\left[\frac{1}{ s \left(\sqrt{s} \coth {\sqrt{s}}-1\right)}\right] \left(\frac{t}{\tau_i}\right) \label{eq:beforetEOD}
\end{eqnarray}
and for $t_{EOD}<t$:
\begin{eqnarray}
&&\theta_i(t)=\Theta_{i,0}\nonumber \\
&&-\zeta_i{\mathcal{L}}^{-1}\left[\frac{1}{ s \left(\sqrt{s} \coth {\sqrt{s}}-1\right)}\right]|^{\left(\frac{t}{\tau_i}\right)}_{\left(\frac{t-t_{EOD}}{\tau_i}\right)}\label{eq:solbSPM3}.
\end{eqnarray}
where we used notation $F(t)|^{x}_{y}\equiv F(x)-F(y)$ for brevity. 

Useful asymptotic expressions for $\theta_i(t)$ can be obtained explicitly for $t \ll \tau_i$  
\begin{eqnarray}
&&\theta_i(t)=\Theta_{i,0}-2\zeta_i\sqrt{\frac{t}{\pi \tau_i}}\label{eq:solas1} 
\end{eqnarray}
and for $\tau_i \ll t \le t_{EOD}$:
\begin{eqnarray}
&&\theta_i(t)=\Theta_{i,0}-\zeta_i \left(\frac{3t}{\tau_i}+\frac{1}{5}\right). \label{eq:solas2} 
\end{eqnarray}

Another useful quantity is the spatially averaged filling factor, which can be calculated from Eq.(\ref{eq:InvLaplaceSPM}):
\begin{eqnarray}
&&\left(\Theta_i\right)_{av}=3\int_0^1 dr \ r^2 \Theta_i(r,t)\nonumber \\
&&=\Theta_{i,0}-3{\mathcal{L}}^{-1}\left[\frac{{\mathcal{L}}\left[\zeta_i(t)\right] }{s}\right](t) \nonumber \\
&&=\Theta_{i,0}-3\int_0^t\zeta_i(t) dt'. \label{eq:InvLaplaceSPM_av}
\end{eqnarray}

\subsection{Voltage dynamics}

For finite discharge rate and under assumptions of the SPM (in the zero order in electrolyte effects) the voltage drop on each electrode equals the sum of the open circuit voltage, electrochemical overpotential and Ohmic drop:
\begin{eqnarray}
&&\Delta \phi_i =\Delta \phi^{eq}_i(\bar{\theta}_i)+\eta_i +I_i r_i, \label{eq:SPM_pot}
\end{eqnarray}
where $I_a=-I_c$, $\left|I_i\right|=I$ and $I_a>0$ for discharging current by convention.  the Ohmic resistance is an empirical parameter, and the overpotential is assumed to follow from the Butler-Volmer kinetics:
\begin{eqnarray}
&&j_i=j_{i,0}(\bar{\theta}_i)\left(e^{\left(1-\alpha_i\right)\frac{e \eta_i}{k_B T}}-e^{-\alpha_i \frac{e \eta_i}{k_B T}}\right), \label{eq:SPM_BV}
\end{eqnarray}
where $j_i$ is the  current density across the SEI and $j_{i,0}$ is the exchange current for the electrode $i=a,c$. We note that choice of the charge transfer coefficient $0 <\alpha_i <1$ is not a simple matter \cite{Guidelli2014}; generalizations of or alternatives to  Butler-Volmer kinetics have also been proposed recently \cite{Bazant2013}. However, a common modeling practice is to take $\alpha_i=1/2$. We shall make this choice in the present work. It allows us to easily invert the expression (\ref{eq:SPM_BV}) for the overpotential:
\begin{eqnarray}
&&\eta_i =\frac{2 k_B T}{e} \ln\left(\chi_i(\bar{\theta}_i)+\sqrt{\chi_i(\bar{\theta}_i)^2+1}\right), \label{eq:SPM_opot}\\
&& \chi_i(\bar{\theta}_i)\equiv \frac{j_i}{j_{i,0}(\bar{\theta}_i)}, \nonumber
\end{eqnarray}
where during discharging $\eta_c, \chi_c <0$ and $\eta_a, \chi_a >0$, and during charging, the other way around. Using expressions (\ref{eq:SPM_pot}) and (\ref{eq:SPM_opot}) we obtain for the terminal voltage:
\begin{eqnarray}
&&V=\Delta \phi^{eq}_c(\bar{\theta}_c)-\Delta \phi^{eq}_a(\bar{\theta}_a)-I_a \left(r_a+r_c\right) \nonumber \\
&&+ \frac{2 k_B T}{e} \ln\left(\frac{\chi_c(\bar{\theta}_c)+\sqrt{\chi_c(\bar{\theta}_c)^2+1}}{\chi_a(\bar{\theta}_a)+\sqrt{\chi_a(\bar{\theta}_a)^2+1}}\right), \label{eq:SPM_tot}
\end{eqnarray}
which, together with Eqs.(\ref{eq:solbSPM}) constitute the zero-order (in the electrolyte and temperature effects) solution for the terminal voltage dynamics.

In the expression (\ref{eq:SPM_tot}) the exchange currents $j_{i,0}$ are yet to be specified. It is customarily to use an excluded-volume generalization of the dilute solution expression: 
\begin{eqnarray}
&&j_{i,0}(\bar{\theta}_i,c_+)=k_i c_+^{1-\alpha} {\theta}_i^{\alpha_i} \left(1-{\theta}_i\right)^{1-\alpha_i},  \label{eq:exchange_current}\\
&& {\theta}_a=\bar{\theta}_a \Theta_{a,0}, \ \  {\theta}_c=1-\bar{\theta}_c\left(1- \Theta_{c,0}\right), \nonumber
\end{eqnarray}
where $c_+$ is the concentration of Li ions in the electrolyte.

%It is seen that $\Theta_i^*$ do affect the dynamics of voltage. Modeling its aging is required.  Based on the previous section and the present one we conclude that modeling aging requires both aging of $\Theta_i^*$ and $M_{Li,i}$, the total number of Li intercalation sites in electrode for $i=a,c$, i.e., four parameters altogether. Eqs.(\ref{eq:solbSPM2}) imply that two additional parameters to model are $\tau_i$.

The resulting explicit expressions for $\chi_i$ are:
\begin{eqnarray}
&&  \chi_i  (\bar{\theta}_i)= \frac{j_i  }{j_{i,0}(\bar{\theta}_i  )}=\frac{I_i R_i}{3 A l_i (1-\epsilon_i)j_{i,0}(\bar{\theta}_i  )}\nonumber \\
&&=\frac{I_i R_i}{3 A l_i (1-\epsilon_i)k_i c_e^{1-\alpha_i} \left({\theta}_i  \right)^{\alpha_i} \left(1-{\theta}_i  \right)^{1-\alpha_i}} \nonumber \\
&&=\frac{I_i}{\tilde{I}_i \left({\theta}_i  \right)^{\alpha_i} \left(1-{\theta}_i  \right)^{1-\alpha_i}}, \label{eq:xhi_explct}\\
&&\tilde{I}_i\equiv  \frac{3 A l_i (1-\epsilon_i)k_i c_e^{1-\alpha_i}}{R_i}, \nonumber \\
&& {\theta}_a=\bar{\theta}_a \Theta_{a,0}, \ \  {\theta}_c=1-\bar{\theta}_c\left(1- \Theta_{c,0}\right). \nonumber
\end{eqnarray}

\section{Fitting the anode and cathode OCPs} \label{app:OCP}
SPM requires open-circuit potentials (OCP) for each electrode, $\Delta \phi^{eq}_c( \bar{\theta}_c)$ and $\Delta \phi^{eq}_a( \bar{\theta}_a)$, Eq.(\ref{eq:governing}), where $\Delta \phi^{eq}_a(x)$ is the anode OCP, corresponding to the state of charge $x$, and $\Delta \phi^{eq}_c(x)$ can be defined as follows:
\begin{eqnarray}
&&\Delta \phi^{eq}_c(x)=U_{OCP}(y)+\Delta \phi^{eq}_a(y), \label{eq:OCV} \\
&& y=1-\frac{N_{Li,c}}{N_{Li,a}}\left(1-x\right),\label{eq:OCV_thetac}
\end{eqnarray}
where  $U_{OCP}(x)$ is the total cell OCP, corresponding to the state of charge $x$, $N_{Li,a}$ the total number of ions available for intercalation in the anode, and $N_{Li,c}$ is the total number of vacancies in the cathode in fully charged cell. 

To obtain the total OCP as a function of the state of charge, the cell voltage $V_{OCP}(t)$ was measured at low discharge current ($0.025C$ for $F1$ case and $0.05C$ for $F3$ case). For such low discharge rate:
\begin{eqnarray}
&&U_{OCP}( \bar{\theta}_a)=U_{OCP}(1-t/t_{OCP}^*), \label{eq:extrap}
\end{eqnarray}
where $t_{OCP}^*$ is the extrapolated time at which $V(t)$ diverges ("absolute" time of the end of discharge). Assuming the divergence is logarithmic, i.e., $V(t) \sim \log(1-t/t_{OCP}^*)$ as $t \rightarrow t_{OCP}^*$, we can numerically perform the extrapolation to find $t^*$ and to obtain the functional form $U_{OCP}(x)$ from Eq.(\ref{eq:extrap}). 
 
Non-destructive measurements access only the total open-circuit voltage. OCPs $\Delta \phi^{eq}_c( \bar{\theta}_c)$ and $\Delta \phi^{eq}_a( \bar{\theta}_a)$ are inaccessible. A common approach for modeling cells with graphite anodes is to assume that the anode OCP is similar to an anode OCP reported in literature and to infer the cathode OCP using \ref{eq:OCV}. The problem with this approach that that the anode OCPs vary significantly and the resulting cathode OCP will generally inherit some features from the assumed graphite anode OCP. 
Therefore we took a different path. We assumed that the anode OCP is described by the model of Ref. \cite{Verbrugge2003} parametrized by  parameters $A_j$, $B_j$ and $C_j$, $j=1,2,...,6$:
\begin{eqnarray}
&&\theta_a=\sum_j\frac{A_j}{1+\exp\left[B_j V+C_j\right]}, \label{eq:Verbrugge}
\end{eqnarray}
where $\theta_a$ is the intercalation fraction (fractional occupancy, \cite{Verbrugge2003}) in the anode at equilibrium, and $V$ is the corresponding anode OCP. We note, based on the definitions in  Eqs.(\ref{eq:theta_norm}), that the corresponding state of charge is $\bar{\theta}_a=\theta_a/\Theta_{a,0}$, where 
\begin{eqnarray}
&&\Theta_{a,0}=\sum_j\frac{A_j}{1+\exp\left[B_j V_c+C_j\right]}, \label{eq:Verbrugge_Vc}
\end{eqnarray}
where $V_c$ is voltage corresponding to the fully charged battery.

Next, we made use of the fact that LiCO2 cathode OCP are smooth \cite{Megahed94} and don't display sharp transitions as observed in both graphite anode and total cell OCPs. Therefore, we assumed that all these features in the total cell OCP should be attributed to the graphite anode. Accordingly, the parameters of the \cite{Verbrugge2003} model we varied to provide the maximally smooth cathode OCP, as shown in Figure \ref{fig:OCP_fit}. The resulting anode and cathode OCP were used in the Eq.(\ref{eq:governing}). 

\section{Definition of the parametric boundaries and other hyperparameters} \label{app:hyperparameters}
Table \ref{table:hyperparameters} lists hyperparameters used to set up the boundaries for the allowed parameters variation. They are based on prior estimates $N_{Li,a}^*$, $\tau_a^*$, $r^*$ and $\Theta_{a}^*$, for the number of ions available for intercalation (the cell capacity), solid phase diffusion time, Ohmic resistance and the maximal filling fraction in anode, respectively.

\begin{table*}[t]
\centering
\caption{Boundaries of the parametric space (hyperparameters).}
\label{table:hyperparameters}
\begin{tabular}{llll}
\hline\hline
  Parameters &\vline \ \ \begin{tabular}[c]{@{}l@{}}Lower\\   boundary\end{tabular} &\vline \ \ \begin{tabular}[c]{@{}l@{}}Upper\\   boundary\end{tabular} &\vline \ \  Remarks  \\ \hline
$\tau_c$  &\vline \ \ $0.001 \tau_a^*$ &\vline \ \  $2 \tau_a^*$ &\vline \ \  $\tau_a^*$ in Appendix \ref{app:hyperparameters} \\ \hline 
$N_{Li,c}$ &\vline \ \ $1.01N_{Li,a}^*$ &\vline \ \  $2N_{Li,a}^*$ &\vline \ \  $N_{Li,a}^*$ in Appendix \ref{app:hyperparameters} \\  \hline
$\Theta_{c,0}$  &\vline \ \ $0$ &\vline \ \  $0.8$ &\vline \ \  $0.5$ is a literature value \\ \hline 
$\tilde{I}_a$  &\vline \ \ $0.1A$ &\vline \ \  $10A$ &\vline \ \  $1A$ is taken as the characteristic scale\\ \hline
$\tilde{I}_c$  &\vline \ \ $0.1A$ &\vline \ \  $10A$ &\vline \ \  $1A$ is taken as the characteristic scale \\ \hline 
$r$   &\vline \ \ $0$ &\vline \ \  $1.2r^*$ &\vline \ \  $r^*$ in Appendix \ref{app:hyperparameters}\\ \hline
$N_{Li,a}$ &\vline \ \ $0.9N_{Li,a}^*$ &\vline \ \  $1.1N_{Li,a}^*$ &\vline \ \  $N_{Li,a}^*$ in Appendix \ref{app:hyperparameters} \\ \hline 
$\tau_a$  &\vline \ \ $0.1 \tau_a^*$ &\vline \ \  $5 \tau_a^*$ &\vline \ \  $\tau_a^*$ in Appendix \ref{app:hyperparameters} \\ \hline
$\Theta_{a,0}$ &\vline \ \ $\Theta_{a}^*$ &\vline \ \  $\Theta_{a}^*$ &\vline \ \  \begin{tabular}[c]{@{}l@{}} $\Theta_{a}^*$ as obtained from the anode OCP \\  model, Appendix \ref{app:OCP}\end{tabular}  \\   
\hline\hline                                                          
\end{tabular}
\end{table*}
\begin{table*}[t]
\centering
\caption{Quality of fits. Calculated at the center of mass of the $\epsilon \le 1.02 \epsilon_{min|tr}$ ensemble.}
\label{table:Q_of_Fits}
\begin{tabular}{lllllll}
\hline\hline
   &\vline \ \ &\vline \ \ 2A &\vline \ \  1.5A &\vline \ \  1A &\vline \ \  0.5A &\vline \ \  0.055A \\ \hline
F1  \ \  &\vline \ \ \begin{tabular}[c]{@{}l@{}}$\epsilon_{t_{EOD}}$\\   $\epsilon_{V(t)}$\end{tabular} &\vline \ \  \begin{tabular}[c]{@{}l@{}}-1.2 \%\\   1.5e-02 V\end{tabular} &\vline \ \  \begin{tabular}[c]{@{}l@{}}0.5 \%\\   2.1e-02 V\end{tabular} &\vline \ \  \begin{tabular}[c]{@{}l@{}}1.8 \%\\   2.9e-02 V\end{tabular} &\vline \ \  \begin{tabular}[c]{@{}l@{}}2 \%\\   3.3e-02 V\end{tabular} &\vline \ \  \begin{tabular}[c]{@{}l@{}}-0.1 \%\\   1.1e-02 V\end{tabular}\\ \hline
F1b &\vline \ \ \begin{tabular}[c]{@{}l@{}}$\epsilon_{t_{EOD}}$\\   $\epsilon_{V(t)}$\end{tabular} &\vline \ \  \begin{tabular}[c]{@{}l@{}}-1.3 \%\\   1.6e-02 V\end{tabular} &\vline \ \  \begin{tabular}[c]{@{}l@{}}0.5 \%\\   2.2e-02 V\end{tabular} &\vline \ \  \begin{tabular}[c]{@{}l@{}}1.8 \%\\   3.1e-02 V\end{tabular} &\vline \ \  \begin{tabular}[c]{@{}l@{}}2 \%\\   3.4e-02 V\end{tabular} &\vline \ \  \begin{tabular}[c]{@{}l@{}}-0.1 \%\\   1.1e-02 V\end{tabular} \\
\hline
F2 &\vline \ \ \begin{tabular}[c]{@{}l@{}}$\epsilon_{t_{EOD}}$\\   $\epsilon_{V(t)}$\end{tabular} &\vline \ \  \begin{tabular}[c]{@{}l@{}}-0.7 \%\\   1.5e-02 V\end{tabular} &\vline \ \  \begin{tabular}[c]{@{}l@{}}1.3 \%\\   2.8e-02 V\end{tabular} &\vline \ \  \begin{tabular}[c]{@{}l@{}}2.6 \%\\   4.2e-02 V\end{tabular} &\vline \ \  \begin{tabular}[c]{@{}l@{}}2.6 \%\\   4.4e-02 V\end{tabular} &\vline \ \  \begin{tabular}[c]{@{}l@{}}0.1 \%\\   2.1e-02 V\end{tabular} \\
\hline\hline  
F2b &\vline \ \ \begin{tabular}[c]{@{}l@{}}$\epsilon_{t_{EOD}}$\\   $\epsilon_{V(t)}$\end{tabular} &\vline \ \  \begin{tabular}[c]{@{}l@{}}-1.2 \%\\   1.5e-02 V\end{tabular} &\vline \ \  \begin{tabular}[c]{@{}l@{}}0.6 \%\\   2.1e-02 V\end{tabular} &\vline \ \  \begin{tabular}[c]{@{}l@{}}1.9 \%\\   3.1 e-02 V\end{tabular} &\vline \ \  \begin{tabular}[c]{@{}l@{}}2.1 \%\\   3.4e-02 V\end{tabular} &\vline \ \  \begin{tabular}[c]{@{}l@{}}-0.1 \%\\   1.0e-02 V\end{tabular} \\
\hline\hline                                                          
\end{tabular}
\end{table*}

\begin{table*}[t]
\centering
\caption{Projections of the best-fit ensemble displacement on the principle axes.}
\label{table:D_of_Shifts}
\begin{tabular}{lllllll}
\hline\hline
   &\vline \ \ $V_1$ &\vline \ \ $V_2$ &\vline \ \  $V_3$ &\vline \ \  $V_4$ &\vline \ \  $V_5$ &\vline \ \  $V_6$ \\ \hline
F1 \ \  &\vline \ \ $-$ &\vline \ \  $-$ &\vline \ \  $-$ &\vline \ \ $-$ &\vline \ \  $-$ &\vline \ \  $-$\\ \hline
F1b &\vline \ \ 0.0000 &\vline \ \      0.0002  &\vline \ \     0.0066 &\vline \ \      0.0728 &\vline \ \      0.0115 &\vline \ \      0.2994  \\
\hline
F2 &\vline \ \ 0.0000  &\vline \ \     0.0001 &\vline \ \     0.0333  &\vline \ \     3.6535 &\vline \ \      0.9818&\vline \ \      1.0672 \\
\hline\hline  
F2b &\vline \ \ 0.0000    &\vline \ \   0.0001   &\vline \ \   0.0004     &\vline \ \   0.0334     &\vline \ \  0.0062    &\vline \ \   0.3347 \\
\hline\hline                                                          
\end{tabular}
\end{table*}

\begin{table*}[tbp]
\centering
\caption{Quality of fits. Calculated at the center of mass of the $\epsilon \le 1.02 \epsilon_{min|tr}$ ensemble.}
\label{table:Q_of_Fits2}
\begin{tabular}{lllllll}
\hline\hline
   &\vline \ \ &\vline \ \ 2A &\vline \ \  1.5A &\vline \ \  $1A/1A^*$  &\vline \ \  0.5A &\vline \ \  0.11A \\ \hline
F3  \ \  &\vline \ \ \begin{tabular}[c]{@{}l@{}}$\epsilon_{t_{EOD}}$\\   $\epsilon_{V(t)}$\end{tabular} &\vline \ \  \begin{tabular}[c]{@{}l@{}}-2.2 \%\\   1.4e-02 V\end{tabular} &\vline \ \  \begin{tabular}[c]{@{}l@{}}$-$  \\   $-$ \end{tabular} &\vline \ \  \begin{tabular}[c]{@{}l@{}}0.5 \%\\   2.3e-02 V\end{tabular} &\vline \ \  \begin{tabular}[c]{@{}l@{}}$-$ \\   $-$ \end{tabular} &\vline \ \  \begin{tabular}[c]{@{}l@{}}-0.1 \%\\   2.4e-02 V\end{tabular}\\ \hline
F4  &\vline \ \ \begin{tabular}[c]{@{}l@{}}$\epsilon_{t_{EOD}}$\\   $\epsilon_{V(t)}$\end{tabular} &\vline \ \  \begin{tabular}[c]{@{}l@{}}-2.2 \%\\   1.4e-02 V\end{tabular} &\vline \ \  \begin{tabular}[c]{@{}l@{}} $-$ \\   $-$ \end{tabular} &\vline \ \  \begin{tabular}[c]{@{}l@{}}0.5/2.3 \%\\   2.3/2.8e-02 V\end{tabular} &\vline \ \  \begin{tabular}[c]{@{}l@{}}$-$ \\   $-$ \end{tabular} &\vline \ \  \begin{tabular}[c]{@{}l@{}}0.2 \%\\   2.6e-02 V\end{tabular} \\
\hline\hline                                                          
\end{tabular}
\end{table*}

\begin{table}[tbp]
\centering
\caption{Projections of the ensemble displacement on the principle axes.}
\label{table:D_of_Shifts2}
\begin{tabular}{lllllll}
\hline\hline
   &\vline \ \ $V_1$ &\vline \ \ $V_2$ &\vline \ \  $V_3$ &\vline \ \  $V_4$ &\vline \ \  $V_5$ &\vline \ \  $V_6$ \\ \hline
F3  \ \  &\vline \ \ $-$ &\vline \ \  $-$ &\vline \ \  $-$ &\vline \ \ $-$ &\vline \ \  $-$ &\vline \ \  $-$\\ \hline
F4  &\vline \ \  0.0000    
 &\vline \ \ 0.0000    
 &\vline \ \  0.0034     &\vline \ \  0.0715 &\vline \ \      0.0166   &\vline \ \   0.0230  \\
\hline\hline                                                          
\end{tabular}
\end{table}

The estimate $N_{Li,a}^*$ is obtained from the measurement of the total cell OCP:
\begin{eqnarray}
&&N_{Li,a}^*=e^{-1} I_{OCP} t_{OCP}^*, \label{eq:extrap2}
\end{eqnarray}
where $I_{OCP}$ is the constant discharge current used for testing the OCP as described in Appendix \ref{app:OCP} and absolute time of the end of discharge $t_{OCP}^*$ is defined after Eq.(\ref{eq:extrap}).

The estimate for the solid phase diffusion time in the anode $\tau_a^*$ is obtained by solving simultaneously the system of two equations:
\begin{eqnarray} 
&&0=1-{\mathcal{L}}^{-1}\left[\frac{\bar{\zeta}_a}{ s \left(\sqrt{s} \coth {\sqrt{s}}-1\right)}\right] \left(\frac{t^*}{\tau_a^*}\right), \label{eq:tau_a_star}\\
&&\bar{\zeta}_a= \frac{I \tau_i^* }{3 e N_{Li,a}^*},
\end{eqnarray}
which follow from Eq.(\ref{eq:solbSPM2_sum}).  In Eq.(\ref{eq:tau_a_star}) current $I \gg I_{OCP}$ and $t^*$ is the corresponding absolute time of discharge.

The characteristic Ohmic resistance $r^*$ is obtained assuming it's the major contribution for the potential drop $\Delta V_I$ with occurs when the discharge at current $I$ is turned on:
\begin{eqnarray} 
&&r^*=\frac{\Delta V_I}{I}, \label{eq:Ohmic}
\end{eqnarray}
The estimate in Eq.(\ref{eq:Ohmic}) was found to weakly depend on the current $I$ and was performed for discharges corresponding to $1C$ or $0.5C$.

Finally, the estimate for maximal filling fraction $\Theta_{a}^*$ in Eq.(\ref{eq:governing}) is given by the expression \ref{eq:Verbrugge} for $V=0$.

\section{Linear analysis of the best-fit ensemble} \label{app:linear}
Here we present a linear analysis of the best-fit ensemble in the vicinity of the minimal error and demonstrate the sensitivity of the best fit to   the perturbation.  As mentioned in Section \ref{sec:sloppy}, one can expect that perturbations in the model, inference algorithm and data will generically lead to large variations of the inferred best-fit values in the sloppy directions, associated with the near-singular eigenvalues of $H^{-1}$. 
Indeed, perturbation in the data $\textbf{x} \rightarrow \textbf{x}+\delta \textbf{x}$ will lead to perturbation of the best fit parameters $\bar{\textbf{p}}^0 \rightarrow \bar{\textbf{p}}^0+\delta \bar{\textbf{p}}$, which can be determined from the variation of Eq.(\ref{eq:minerror}):
\begin{eqnarray}
&&\frac{\partial C(\textbf{x}+\delta \textbf{x},\bar{\textbf{p}})}{\partial \bar{p}_j}|_{\bar{\textbf{p}}^0+\delta \bar{\textbf{p}}}=0  \nonumber \\
&&\Rightarrow \frac{\partial^2 C(\textbf{x},\bar{\textbf{p}})}{\partial \bar{p}_j \partial x_k}|_{\bar{\textbf{p}}^0}\delta x_k+\frac{\partial^2 C(\textbf{x},\bar{\textbf{p}})}{\partial \bar{p}_j \partial \bar{p}_k}|_{\bar{\textbf{p}}^0}\delta \bar{p}_k=0 \nonumber \\
&&\Rightarrow \delta \bar{p}_k=-H^{-1}_{k,j}\frac{\partial^2 C(\textbf{x},\bar{\textbf{p}})}{\partial \bar{p}_j \partial x_l}|_{\bar{\textbf{p}}^0}\delta x_l. \label{eq:minerror_var1}
\end{eqnarray}
Since the inverse Hessian $H^{-1}_{k,j}$ is ill-conditioned, small generic perturbations in the data $\delta \textbf{x}$ will lead to large variations of the best-fit parametric values $\delta \bar{\textbf{p}}$ in the sloppy directions associated with the near-zero values of the Hessian.

Perturbation in the model $C \rightarrow C+\delta C$ will lead to perturbation of the best fit parameters $\bar{\textbf{p}}^0 \rightarrow \bar{\textbf{p}}^0+\delta \bar{\textbf{p}}$, which can similarly be determined from the variation of Eq.(\ref{eq:minerror}):
\begin{eqnarray}
&&\frac{\partial C(\textbf{x},\bar{\textbf{p}})+\partial \delta C(\textbf{x},\bar{\textbf{p}})}{\partial \bar{p}_j}|_{\bar{\textbf{p}}^0+\delta \bar{\textbf{p}}}=0  \nonumber \\
&&\Rightarrow \frac{\partial^2 C(\textbf{x},\bar{\textbf{p}})}{\partial \bar{p}_j \partial \bar{p}_k}|_{\bar{\textbf{p}}^0}\delta \bar{p}_k+\frac{\partial \delta C(\textbf{x},\bar{\textbf{p}})}{\partial \bar{p}_j}|_{\bar{\textbf{p}}^0}=0 \nonumber \\
&&\Rightarrow \delta \bar{p}_k=-H^{-1}_{k,j}\frac{\partial \delta C(\textbf{x},\bar{\textbf{p}})}{\partial \bar{p}_j}|_{\bar{\textbf{p}}^0}.\label{eq:minerror_var2}
\end{eqnarray}
Again, due to the large condition number of the inverse Hessian $H^{-1}_{k,j}$, small  perturbations of the model $\delta C$ will lead to large variations of the best-fit parametric values $\delta \bar{\textbf{p}}$.

In a practical implementation of a least squares minimization algorithm a stopping criterion is used. Assuming the stopping criterion is satisfied at a point $\bar{\textbf{p}}^0+\delta \bar{\textbf{p}}$, where $\partial C/\partial \bar{p}_j \le \delta g_j$, we can calculate the best-fit value deviation $\delta \bar{\textbf{p}}$ as follows: 
\begin{eqnarray}
&&\frac{\partial C(\textbf{x},\bar{\textbf{p}})}{\partial \bar{p}_j}|_{\bar{\textbf{p}}^0+\delta \bar{\textbf{p}}}=\delta g_j  \nonumber \\
&&\Rightarrow \frac{\partial^2 C(\textbf{x},\bar{\textbf{p}})}{\partial \bar{p}_j \partial \bar{p}_k}|_{\bar{\textbf{p}}^0}\delta \bar{p}_k=\delta g_j \nonumber \\
&&\Rightarrow \delta \bar{p}_k=H^{-1}_{k,j}\delta g_j. \label{eq:minerror_var3}
\end{eqnarray}
Even if the stopping criterion corresponds to a small gradient $\delta \textbf{g}$ the variation of the best-fit parametric values $\delta \bar{\textbf{p}}$ will generically be large due to the ill-conditioning of the inverse Hessian.

Below we explore this predicted effect of perturbations numerically for the SPM. We shall see that the best fit changes significantly in the sloppy directions under the small perturbations while the values of stiff parametric combinations are only negligibly affected.  The following three classes of perturbations are considered:

\begin{itemize}
\item{Perturbations of parametric boundaries (termed cases $F1$ vs. $F1b$, for brevity)}
\item{Variation in the choice of the training data set (cases $F1$ vs. $F2$ and  $F2b$)}
\item{Small variations in the experimental conditions  (cases $F3$ vs. $F4$)}
\end{itemize}

For each perturbation $F_i$ an ensemble of nondimensionalized parametric values corresponding to the RMSE $\epsilon \le 1.02 \epsilon_{min|tr}$is computed and it's center-of-mass value $\mathbf{\bar{p}}_{F_i}$ calculated. Comparing cases $F_i$ and $F_j$ we compute the displacement vector $\Delta \mathbf{\bar{p}}|^{F_j}_{F_i}\equiv \mathbf{\bar{p}}_{F_j}-\mathbf{\bar{p}}_{F_i}$ and it's projections on the PAs of ensemble, corresponding to $F_i$ (without loss of generality, as the PAs are only weakly perturbed), Tables \ref{table:D_of_Shifts} and \ref{table:D_of_Shifts2}. Tables \ref{table:Q_of_Fits} and \ref{table:Q_of_Fits2} compare the qualities of the fit to constant discharge data in terms of the RMSE, associated with cases $F_i$ and $F_j$ 

\paragraph{Perturbation of parametric boundaries} We have considered perturbation the boundary of parameter $\Theta_c^*$: the upper bounds for $\Theta_{c,0}$ is set to $0.4$ in $F1b$ instead of $0.8$ in $F1$. Table \ref{table:Q_of_Fits} shows that the quality of the fit is unchanged by the perturbation.

\paragraph{Perturbation of the training data set} 
Best-fit ensembles are compared for fits $F1$ vs. $F2$. Fit $F2$ differs from $F1$ in that the discharge data for $1A$ is excluded from the training set. Again, the  ensembles are found to have shifted in the sloppy directions only, Table \ref{table:D_of_Shifts}.  The quality of the fit has decreased insignificantly, Table \ref{table:Q_of_Fits}. Case $F2b$ correspond to a different variation in the choice of the training data set, where $1.5A$ discharge data is used for the fitting instead of $1A$ discharge (case $F1$). Again, the  ensembles are found to have shifted in the sloppy directions only, Table \ref{table:D_of_Shifts}.  The qualities of the fits  are practically identical, Table \ref{table:Q_of_Fits}.

\paragraph{Perturbation of experimental conditions} 
Best-fit ensembles are compared for fits $F3$ vs. $F4$ for a series of tests on moderately aged cells, where such data was available. Fit $F4$ differs from $F3$ in that the discharge data for $1A$ (starred value in Table \ref{table:Q_of_Fits2})  is obtained in a different although nominally identical test. Again, the ensembles are found to have shifted in the sloppy directions only, Table \ref{table:D_of_Shifts2}.  The quality of the fit has decreased insignificantly, Table \ref{table:Q_of_Fits2}.

The foregoing analysis illustrates an expected property of "sloppy models": small (in the RMSE metric) perturbations of the parametric boundaries, training data set and experimental conditions, lead to significant variations of the best fit values of the parameters. These significant variations for all the perturbations take place in particular (sloppy) directions in the parametric space. Variations in the perpendicular directions (stiff) are small. The ratios of the variation magnitudes between the sloppy and stiff directions span many orders of magnitude. The numerical evidence presented above suggests that $\textbf{V}_i$, $i=1,2,3$ are essentially stiff under the considered perturbations, while $\textbf{V}_i$, $i=4,5,6$ are sloppy. Based on the Table \ref{table:PAs_F1_vs_F3b} we find that two out of the three stiff parameters, corresponding to $\textbf{V}_1$ ans $\textbf{V}_2$, are the original ("microscopic") parameters of the SPM: the number of Li ions $N_{Li,a}$  and the diffusion time $\tau_a$ in anode, respectively. The third effective stiff parameter $\tilde{r}$, corresponding to $\textbf{V}_3$  is found to be a combination of the total Ohmic resistance $r$  and the electrodes specific exchange currents $\tilde{I}_i$, $i=a,c$. The quality of the fit, as judged by the accuracy of the predicted time-of end of discharge and the mean square distance from the discharge voltage data, Tables \ref{table:Q_of_Fits} and \ref{table:Q_of_Fits2}, is negligibly perturbed by the considered perturbations, while the variations in the sloppy directions are very significant. This picture suggests a operational definition of \textit{the} stiff directions  as \textit{directions which are stiff under typical perturbations of the experimental conditions and the fitting procedure, which negligibly affects the fitting quality}.

\end{document}